\documentclass[12pt]{article}

\usepackage{amstext}

\def\square{\kern1pt\vbox{\hrule height 1.2pt\hbox{\vrule width 1.2pt\hskip 3pt
   \vbox{\vskip 6pt}\hskip 3pt\vrule width 0.6pt}\hrule height 0.6pt}\kern1pt}

\begin{document}

\begin{titlepage}

\begin{flushright}
UFIFT-QG-12-01
\end{flushright}

\vspace{1.5cm}

\begin{center}
{\bf Linearized Weyl-Weyl Correlator in a de Sitter Breaking Gauge}
\end{center}

\vspace{.5cm}

\begin{center}
P. J. Mora$^{\dagger}$ and R. P. Woodard$^{\ddagger}$
\end{center}

\vspace{.5cm}

\begin{center}
\it{Department of Physics \\
University of Florida \\
Gainesville, FL 32611}
\end{center}

\vspace{1cm}

\begin{center}
ABSTRACT
\end{center}
We use a de Sitter breaking graviton propagator \cite{TW1,RPW1} to
compute the tree order correlator between noncoincident Weyl tensors
on a locally de Sitter background. An explicit, and very simple
result is obtained, for any spacetime dimension $D$, in terms of a
de Sitter invariant length function and the tensor basis constructed
from the metric and derivatives of this length function. Our answer
does not agree with the one derived previously by Kouris
\cite{Kouris}, but that result must be incorrect because it not
transverse and lacks some of the algebraic symmetries of the Weyl
tensor. Taking the coincidence limit of our result (with dimensional
regularization) and contracting the indices gives the expectation
value of the square of the Weyl tensor at lowest order. We propose
the next order computation of this as a true test of de Sitter
invariance in quantum gravity.

\vspace{.5cm}

\begin{flushleft}
\end{flushleft}

\vspace{1.5cm}
\begin{flushleft}
$^{\dagger}$ e-mail: pmora@phys.ufl.edu \\
$^{\ddagger}$ e-mail: woodard@phys.ufl.edu
\end{flushleft}
\end{titlepage}

\section{Introduction}

Students of quantum mechanics are familiar with the fact that
charged particle wave functions couple to the electromagnetic
vector potential, not to the field strength tensor. Hence the
undifferentiated vector potential in a fixed gauge is, in some
ways, observable. This point was crushingly demonstrated by
the famous Aharonov-Bohm effect in which a charged particle is
made to interfere with itself in passing round a solenoid,
despite the field strength being zero throughout the support
of the particle's wave function \cite{bohm}.

Specialists in quantum field theory on curved space are engaged in
a similar debate concerning inflationary gravitons. Matter fields
couple to the metric, not to the curvature. There is no gauge in
which this can be avoided. Hence one would think it obvious that
that the undifferentiated graviton field in a fixed gauge must be
observable. Indeed, strenuous efforts \cite{PLANCK,EBEX,SPIDER,PIPER}
are under way to measure the tensor power spectrum, which is the
expectation value of the conformally rescaled graviton field in
transverse-traceless and synchronous gauge, taken long after the
time $t_k$ of first horizon crossing,
\begin{equation}
\Delta^2_h(k) \equiv \frac{k^3}{2 \pi^2} \lim_{t \gg t_k} \int \!\! d^3x \,
e^{-i \vec{k} \cdot \vec{x}} \Bigl\langle \Omega \Bigl\vert
h^{tt}_{ij}(t,\vec{x}) h^{tt}_{ij}(t,\vec{0}) \Bigr\vert
\Omega \Bigr\rangle \; . \label{Deltah}
\end{equation}

Mathematical physicists have for years disputed this conclusion
because it conflicts with their belief in the de Sitter invariance
of free gravitons on de Sitter background. (The de Sitter geometry
is the most highly accelerated inflation consistent with classical
stability.) The Bunch-Davies mode sum for the graviton propagator is
formally de Sitter invariant, but infrared divergent. Regulating the
infrared divergence breaks de Sitter invariance \cite{TW2}. However,
the infrared divergence is only logarithmic, so the derivatives
needed to turn a graviton field into a linearized curvature render
the mode sum for the linearized Weyl-Weyl correlator infrared finite
and de Sitter invariant. Mathematical physicists therefore find it
attractive to argue that the graviton propagator is unobservable ---
in spite of current efforts \cite{PLANCK,EBEX,SPIDER,PIPER} to
observe tensor power spectrum (\ref{Deltah}) --- and insist that
only the correlator of two linearized Weyl tensors is physical. They
sometimes even advance the de Sitter invariance of the Weyl-Weyl
correlator as evidence that free gravitons are physically de Sitter
invariant \cite{FH,HMM,MTW1}.

A digression is necessary at this stage to mention two recent
insights which have dispelled decades of confusion:
\begin{itemize}
\item{There is a topological obstacle that precludes adding invariant
gauge fixing terms to the action on any manifold, such as de Sitter,
which possesses a linearization instability \cite{MTW2}; and}
\item{It is incorrect to subtract off power law infrared divergences,
which is what automatically happens with any analytic regularization
technique, such as continuation from Euclidean de Sitter space
\cite{MTW3}.}
\end{itemize}
The first point explains that there is no math error, but rather a
subtle physics problem with gauge fixing in the many solutions which
have been reported for the graviton propagator with a covariant gauge
fixing term \cite{INVPROP}. Attempting to ignore this problem produces
provably wrong results in scalar quantum electrodynamics \cite{KW1},
and would do so as well in quantum gravity.

It is still possible to add noncovariant gauge fixing terms to the
action, or to impose a covariant gauge exactly (as opposed to on the
average with a gauge fixing term). The propagator was long ago
worked out with a noncovariant gauge fixing term \cite{TW1,RPW1},
and all quantum gravitational loop corrections on de Sitter have
been made using this solution \cite{TW3,TW4,TW5,MW1,KW2}. Enhancing
the naive de Sitter transformation with the compensating gauge
transformation needed to restore the noncovariant gauge condition
reveals a physical breaking of de Sitter invariance \cite{Kleppe}.
The propagator has also recently been constructed in a covariant,
exact gauge \cite{MTW4}, and that solution shows explicit breaking
of de Sitter invariance as well \cite{KMW}.

The second point of our digression explains the curious statement in the
mathematical literature that exact covariant gauges are free of infrared
problems except for certain discrete values of the gauge fixing parameters
\cite{Higuchi}. It has even been asserted that minimally coupled scalars
with tachyonic masses are infrared finite except for the discrete values,
$M^2 = -N(N+3) H^2$, where $H$ is the Hubble parameter \cite{FH}. In fact,
all tachyonic masses produce infrared divergences. The special thing about
the discrete values is that one of the power law infrared divergences
happens to become logarithmic for these values, and so is not
automatically subtracted by the analytic regularization
technique.\footnote{Mathematical physicists occasionally ask what is
wrong with the de Sitter invariant solutions one gets from subtracting
off power law infrared divergences. The result is a solution to the
propagator equation which is not a propagator in the sense of being the
expectation value of the time-ordered product of two field in the
presence of any normalizable state. Such solutions abound, for example,
$i/2$ times the sum of the advanced and retarded Green's functions
\cite{TW6}.}

We come now to the main point of this paper, which is to evaluate
the linearized Weyl-Weyl correlator in the same noncovariant gauge
\cite{TW1,RPW1} for which all existing quantum gravitational loop
corrections on de Sitter background have been made
\cite{TW3,TW4,TW5,MW1,KW2}. We will demonstrate four things:
\begin{itemize}
\item{That our result is both de Sitter invariant and very simple;}
\item{That the result obtained in 2001 by Kouris \cite{Kouris}
cannot be correct because it possesses neither the algebraic
symmetries of the Weyl tensor, nor its transversality;}
\item{That the de Sitter invariance of our result is a trivial
consequence of the derivatives needed to convert the graviton field
into a linearized curvature and the disappearance of the constrained
parts of the propagator from the linearized Weyl-Weyl correlator;
and}
\item{That a true test of de Sitter invariance lies in
evaluating the next loop order result for the coincident Weyl-Weyl
correlator with its indices properly contracted.}
\end{itemize}
Section 2 deals with the apparatus of perturbative quantum gravity
on a $D$-dimensional de Sitter background so that dimensional
regularization can be used. The actual computation is performed in
section 3. We also discuss the discrepancy between the earlier
result \cite{Kouris} and ours. In section 4 we explain what the
Weyl-Weyl correlator tells one and what it does not. We also compare
it to the expectation value of the stress tensor of a massless,
minimally coupled scalar, both at the free level (which produces a
de Sitter invariant result) and with a quartic self-interaction
(which shows de Sitter breaking).

\section{Quantum Field Theory on de Sitter}

The purpose of this section is to describe the formalism for making
perturbative quantum gravity computations on de Sitter background.
We begin with the open conformal coordinate system which must be
used if de Sitter is to fit into the larger context of inflationary
cosmology. We then present the graviton propagator in our
noncovariant gauge \cite{TW1,RPW1}. The section closes with a
discussion of the tensor basis employed to express the linearized
Weyl-Weyl correlator in a manifestly de Sitter invariant form.

\subsection{Open Conformal Coordinates}

We view de Sitter from the perspective of inflationary cosmology, as
but a special case of the much larger class of homogeneous,
isotropic and spatially flat geometries. This means we do not want
to work on the full de Sitter manifold but rather on the so-called
``cosmological patch'' which is spatially flat. It is convenient to
use conformal coordinates $x^{\mu} = (\eta,\vec{x})$ with,
\begin{equation}
-\infty < \eta < 0 \qquad , \qquad -\infty < x^i < +\infty \qquad
{\rm for} \qquad i = 1, \ldots, D\!-\!1 \; .
\end{equation}
As the name suggests, the metric in these coordinates is conformal
to that of flat space,
\begin{equation}
ds^2 = a^2 \Bigl( -d\eta^2 + d\vec{x} \!\cdot\! d\vec{x} \Bigr)
\qquad {\rm where} \qquad a \equiv -\frac1{H \eta} \; .
\end{equation}
The parameter $H$ is known as the Hubble constant, and is related
to the cosmological constant by $\Lambda = (D-1) H^2$. Although
conformal coordinates do not cover the full de Sitter manifold,
$\eta = {\rm constant}$ does represent a Cauchy surface, so information
from the larger manifold can only enter the cosmological patch as
initial value data.

The symmetry group of coordinate transformations which preserve the
de Sitter metric plays a central role in our analysis. In open
$D$-dimensional conformal coordinates the de Sitter group consists
of $\frac12 D (D + 1)$ transformations which can be arranged as
follows in four parts:
\begin{enumerate}
\item{{\it Spatial Translations}, which comprise $(D - 1)$
transformations parameterized by a constant vector $\epsilon^i$,
\begin{equation}
\eta' = \eta \qquad , \qquad {x'}^i = x^i + \epsilon^i \; .
\label{spacetrans}
\end{equation}}
\item{{\it Spatial Rotations}, which comprise $\frac12 (D-1)(D-2)$
transformations parameterized by the rotation matrix $R^{ij}$,
\begin{equation}
\eta' = \eta \qquad , \qquad {x'}^i = R^{ij} x^j \; .
\label{spacerots}
\end{equation}}
\item{{\it Dilatations}, which comprise one transformation
parameterized by a constant $C$,
\begin{equation}
\eta' = C \eta \qquad , \qquad {x'}^i = C x^i \; .
\label{dilatations}
\end{equation}}
\item{{\it Spatial Special Conformal Transformations}, which
comprise $(D-1)$ transformations parameterized by the constant
vector $\theta^i$,
\begin{equation}
\eta' = \frac{\eta}{1 \!-\! 2 \vec{\theta} \!\cdot\! \vec{x} \!+\!
\theta^2 x^{\mu} x_{\mu} } \qquad , \qquad {x'}^i = \frac{x^i \!-\!
\theta^i x^{\mu} x_{\mu}}{1 \!-\! 2 \vec{\theta} \!\cdot\! \vec{x}
\!+\! \theta^2 x^{\mu} x_{\mu} } \; . \label{spacespec}
\end{equation}}
\end{enumerate}
The symmetries of cosmology are 1 and 2; symmetries 3 and 4 only
appear in the de Sitter limit of maximal acceleration.

It is convenient to represent de Sitter invariant propagators
between points $x^{\mu}$ and ${x'}^{\mu}$ using the de Sitter length
function $y(x;x')$,
\begin{equation}
y(x;x') \equiv a a' H^2 \Biggl[\Bigl\Vert \vec{x} \!-\! \vec{x}'
\Vert^2 - \Bigl( \vert \eta \!-\! \eta' \vert \!-\! i \varepsilon
\Bigr)^2 \Biggr] \; . \label{ydef}
\end{equation}
Except for the factor of $i \varepsilon$ (whose purpose is to
enforce Feynman boundary conditions) the de Sitter length function
can be expressed as follow in terms of the geodesic length
$\ell(x;x')$ from $x^{\mu}$ to ${x'}^{\mu}$,
\begin{equation}
y(x;x') = 4 \sin^2\Bigl( \frac12 H \ell(x;x')\Bigr) \; .
\end{equation}
We should mention that mathematical physicists prefer a different de
Sitter function $z = 1 -\frac14 y$, because it gives simpler
formulae for propagators in terms of hypergeometric functions. The
advantage of our length function $y(x;x')$ is that it vanishes at
coincidence (that is, $x^{\mu} = {x'}^{\mu}$), which is quite
important when renormalizing explicit loop computations.

\subsection{The Graviton Propagator}

We define the graviton field $h_{\mu\nu}(x)$ by conformally
transforming the full metric $g_{\mu\nu}(x)$ and then subtracting
off the background,
\begin{equation}\label{conf}
g_{\mu\nu}(x) \equiv a^2 \widetilde{g}_{\mu\nu} \equiv a^2 \Bigl(
\eta_{\mu\nu} + \kappa h_{\mu\nu}(x)\Bigr) \; . \label{hdef}
\end{equation}
Here $\eta_{\mu\nu}$ is the $D$-dimensional, spacelike signature
Minkowski metric, and $\kappa^2 \equiv 16 \pi G$ is the loop
counting parameter of quantum gravity. The gravitational Lagrangian
is,
\begin{equation}
\mathcal{L} \equiv \frac1{16\pi G} \Bigl( R - (D\!-\!2)
\Lambda\Bigr) \sqrt{-g} \; . \label{Einstein}
\end{equation}
Subtracting off a surface term and expanding in powers of the
graviton field gives a form from which the perturbative interactions
can be read off \cite{TW1},
\begin{eqnarray}
\lefteqn{\mathcal{L} - {\rm Surface} = \Bigl(\frac{D}2 \!-\! 1\Bigr)
H a^{D-1} \sqrt{-\widetilde{g}} \, \widetilde{g}^{\rho\sigma}
\widetilde{g}^{\mu \nu} h_{\rho\sigma ,\mu} h_{\nu 0} + a^{D-2}
\sqrt{-\widetilde{g}} \, \widetilde{g}^{\alpha\beta}
\widetilde{g}^{\rho\sigma} \widetilde{g}^{\mu\nu} } \nonumber \\
& & \hspace{2cm} \times \Biggl\{ \frac12 h_{\alpha\rho ,\mu}
h_{\nu\sigma ,\beta} \!-\! \frac12 h_{\alpha\beta ,\rho}
h_{\sigma\mu ,\nu} \!+\! \frac14 h_{\alpha\beta ,\rho}
h_{\mu\nu ,\sigma} \!-\! \frac14 h_{\alpha\rho ,\mu}
h_{\beta\sigma ,\nu} \Biggr\} . \qquad \label{Linv}
\end{eqnarray}
Note that $\widetilde{g}^{\mu\nu}$ and $\sqrt{-\widetilde{g}}$ are
infinite order in the graviton field,
\begin{eqnarray}
\widetilde{g}^{\mu\nu} & = & \eta^{\mu\nu} - \kappa h^{\mu\nu} + \kappa^2
h^{\mu\rho} h_{\rho}^{~\nu} + O(\kappa^3) \; , \\
\sqrt{-\widetilde{g}} & = & 1 + \frac12 \kappa h + \frac18 \kappa^2 h^2 -
\frac14 \kappa^2 h^{\mu\nu} h_{\mu\nu} + O(\kappa^3) \; .
\end{eqnarray}
Note also that we follow the usual conventions whereby a comma denotes
ordinary differentiation, $h \equiv \eta^{\mu\nu} h_{\mu\nu}$, and
graviton indices are raised and lowered using the Minkowski metric,
$h^{\mu}_{~\nu} \equiv \eta^{\mu\rho} h_{\rho\nu}$ and $h^{\mu\nu}
\equiv \eta^{\mu\rho} \eta^{\nu\sigma} h_{\rho \sigma}$.

The quadratic part of the invariant Lagrangian is,
\begin{equation}
\mathcal{L}^{(2)}_{\rm inv} = \Biggl[ \frac12 h^{\rho\sigma , \mu}
h_{\mu \sigma , \rho} \!-\! \frac12 h^{\mu \nu}_{~~ ,\mu} h_{,\nu}
\!+\! h^{,\mu} h_{, \mu} \!-\! \frac14 h^{\rho\sigma , \mu}
h_{\rho \sigma , \mu} \!-\! \Bigl( \frac{D \!-\! 2}2\Bigr) H a h^{0
\mu} h_{, \mu} \Biggr] a^{D-2} \; .
\end{equation}
To this we add the noncovariant gauge fixing term,
\begin{equation}
\mathcal{L}_{GF} = -\frac12 a^{D-2} \eta^{\mu\nu} F_{\mu} F_{\nu} \; , \;
F_{\mu} \equiv \eta^{\rho\sigma} \Bigl(h_{\mu\rho , \sigma}
- \frac12 h_{\rho \sigma , \mu} + (D \!-\! 2) H a h_{\mu \rho}
\delta^0_{\sigma} \Bigr) .
\end{equation}
Note that it respects de Sitter symmetries 1-3, breaking only
the spatial special conformal transformations. Because space and
time are treated differently in our coordinate system and gauge,
it is useful to have an expression for the purely spatial parts
of the Lorentz metric and the Kronecker delta,
\begin{equation}
\overline{\eta}_{\mu\nu} \equiv \eta_{\mu\nu} + \delta^0_{\mu} \delta^0_{\nu}
\qquad {\rm and} \qquad \overline{\delta}^{\mu}_{\nu} \equiv \delta^{\mu}_{\nu}
- \delta_0^{\mu} \delta^0_{\nu} \; .
\end{equation}
The quadratic part of gauge fixed Lagrangian can be partially integrated
to take the form $\frac12 h^{\mu\nu} D_{\mu\nu}^{~~\rho \sigma}
h_{\rho\sigma}$, where the kinetic operator is,
\begin{eqnarray}
\lefteqn{D_{\mu\nu}^{~~\rho\sigma} \equiv \left\{ \frac12 \overline{\delta}_{
\mu}^{~(\rho} \overline{\delta}_{\nu}^{~\sigma)} - \frac14 \eta_{\mu\nu}
\eta^{\rho\sigma} - \frac1{2(D\!-\!3)} \delta_{\mu}^0 \delta_{\nu}^0
\delta_0^{\rho} \delta_0^{\sigma} \right\} D_A } \nonumber \\
& & \hspace{5cm} + \delta^0_{(\mu} \overline{\delta}_{\nu)}^{(\rho}
\delta_0^{\sigma)} \, D_B + \frac12 \Bigl(\frac{D\!-\!2}{D\!-\!3}\Bigr)
\delta_{\mu}^0 \delta_{\nu}^0 \delta_0^{\rho} \delta_0^{\sigma} \, D_C
\; , \qquad
\end{eqnarray}
and the three scalar differential operators are,
\begin{eqnarray}
D_A & \equiv & \partial_{\mu} \Bigl(a^{D-2} \eta^{\mu\nu}
\partial_{\nu}\Bigr) \; , \\
D_B & \equiv & \partial_{\mu} \Bigl(a^{D-2} \eta^{\mu\nu}
\partial_{\nu}\Bigr) - (D\!-\!2) H^2 a^D \; , \\
D_C & \equiv & \partial_{\mu} \Bigl(a^{D-2} \eta^{\mu\nu}
\partial_{\nu}\Bigr) - 2 (D\!-\!3) H^2 a^D \; .
\end{eqnarray}

The graviton propagator in our gauge takes the form of a sum of
constant index factors times scalar propagators \cite{TW1,RPW1},
\begin{equation}\label{gp}
i\Bigl[{}_{\mu\nu} \Delta_{\rho\sigma}\Bigr](x;x') = \sum_{I=A,B,C}
\Bigl[{}_{\mu\nu} T^I_{\rho\sigma}\Bigr] i\Delta_I(x;x') \; . \label{gprop}
\end{equation}
The three scalar propagators invert the various scalar kinetic operators,
\begin{equation}
D_I \times i\Delta_I(x;x') = i \delta^D(x - x') \qquad {\rm for} \qquad
I = A,B,C \; , \label{sprops}
\end{equation}
and we will give explicit expressions for them. The index factors are,
\begin{eqnarray}
\Bigl[{}_{\mu\nu} T^A_{\rho\sigma}\Bigr] & = & 2 \, \overline{\eta}_{\mu (\rho}
\overline{\eta}_{\sigma) \nu} - \frac2{D\!-\! 3} \overline{\eta}_{\mu\nu}
\overline{\eta}_{\rho \sigma} \; , \label{TA} \\
\Bigl[{}_{\mu\nu} T^B_{\rho\sigma}\Bigr] & = & -4 \delta^0_{(\mu}
\overline{\eta}_{\nu) (\rho} \delta^0_{\sigma)} \; , \label{TB} \\
\Bigl[{}_{\mu\nu} T^C_{\rho\sigma}\Bigr] & = & \frac2{(D \!-\!2) (D \!-\!3)}
\Bigl[(D \!-\!3) \delta^0_{\mu} \delta^0_{\nu} + \overline{\eta}_{\mu\nu}\Bigr]
\Bigl[(D \!-\!3) \delta^0_{\rho} \delta^0_{\sigma} + \overline{\eta}_{\rho
\sigma}\Bigr] \; . \label{TC}
\end{eqnarray}
It is straightforward to verify that the graviton propagator (\ref{gprop})
indeed inverts the gauge-fixed kinetic operator,
\begin{equation}
D_{\mu\nu}^{~~\rho\sigma} \times i\Bigl[{}_{\rho\sigma} \Delta^{\alpha\beta}
\Bigr](x;x') = \delta_{\mu}^{(\alpha} \delta_{\nu}^{\beta)} i \delta^D(x-x')
\; .
\end{equation}

The $A$-type propagator obeys the same equation as that of a massless,
minimally coupled scalar. It has long been known that no de Sitter invariant
solution exists \cite{AF}. If one elects to break de Sitter invariance
while preserving homogeneity and isotropy --- this is known as the ``E(3)''
vacuum \cite{BA} --- the solution takes the form \cite{OW},
\begin{equation}
i\Delta_A(x;x') = A\Bigl( y(x;x') \Bigr) + k \ln(a a') \; , \label{DA}
\end{equation}
where the constant $k$ is,
\begin{equation}
k \equiv \frac{ H^{D-2}}{(4 \pi)^{\frac{D}2}} \frac{\Gamma(D \!-\! 1)}{
\Gamma(\frac{D}2)} \; .
\end{equation}
The function $A(y)$ is,
\begin{eqnarray}\label{Af}
\lefteqn{A(y) = \frac{H^{D-2}}{(4\pi)^{\frac{D}2}} \Biggl\{
\Gamma\Bigl(\frac{D}2 \!-\!1\Bigr) \Bigl(\frac{4}{y}\Bigr)^{
\frac{D}2 -1} \!+\! \frac{\Gamma(\frac{D}2 \!+\! 1)}{\frac{D}2
\!-\! 2} \Bigl(\frac{4}{y} \Bigr)^{\frac{D}2-2} \!+\! A_1 }
\nonumber \\
& & \hspace{1.5cm} - \sum_{n=1}^{\infty} \Biggl[
\frac{\Gamma(n\!+\!\frac{D}2\!+\!1)}{(n\!-\!\frac{D}2\!+\!2) (n
\!+\! 1)!} \Bigl(\frac{y}4 \Bigr)^{n - \frac{D}2 +2} \!\!\!\!\! -
\frac{\Gamma(n \!+\! D \!-\! 1)}{n \Gamma(n \!+\! \frac{D}2)}
\Bigl(\frac{y}4 \Bigr)^n \Biggr] \Biggr\} , \qquad \label{DeltaA}
\end{eqnarray}
where the constant $A_1$ is,
\begin{equation}
A_1 = \frac{\Gamma(D\!-\!1)}{\Gamma(\frac{D}2)} \Biggl\{
-\psi\Bigl(1 \!-\! \frac{D}2\Bigr) + \psi\Bigl(\frac{D\!-\!1}2\Bigr)
+
\psi(D \!-\!1) + \psi(1) \Biggr\} .
\end{equation}
It should be noted that $A(y)$ obeys the differential equation,
\begin{equation}\label{ape}
(4 y \!-\! y^2) A''(y) + D (2 \!-\! y) A'(y) = (D \!-\! 1) k \; .
\end{equation}

The $B$-type and $C$-type propagators are both de Sitter invariant,
\begin{equation}
i\Delta_B(x;x') = B\Bigl(y(x;x') \Bigr) \qquad , \qquad
i\Delta_C(x;x') = C\Bigl(y(x;x')\Bigr) \; .
\end{equation}
Rather than give the series expansion for $B(y)$ we present its
relation to $A(y)$ \cite{MTW1},
\begin{equation}
B(y) = - \frac{[(4y \!-\! y^2) A'(y) \!+\! k (2 \!-\! y)]}{2 (D
\!-\! 2)} \; .
\end{equation}
For $C(y)$ it is more convenient to give the derivative \cite{MTW1},
\begin{equation}
C'(y) = A'(y) - \frac14 \Bigl( \frac{D \!-\! 3}{D \!-\! 2}\Bigr)
\Bigl[ (4y \!-\! y^2) A'(y) \!+\! k (2 \!-\! y) \Bigr] \; .
\end{equation}

Of course our propagator breaks the 4th part of the de Sitter group
(spatial special conformal transformations) because the gauge condition
breaks it. However, the propagator also breaks the 3rd part of the
de Sitter group (dilatations), which is preserved by the gauge condition.
This is evident from the de Sitter breaking second term of the $A$-type
propagator (\ref{DA}), which is needed to reproduce the famous result
for the coincidence limit of the massless, minimally coupled scalar
propagator \cite{VFLS},
\begin{equation}
\lim_{x \rightarrow x'} i\Delta_A(x;x') = \frac{H^2}{4 \pi^2} \, \ln(a)
+ {\rm Divergent\ Constant} \; .
\end{equation}
The absence of dilatation invariance implies a physical breaking of
de Sitter invariance by free gravitons. Kleppe proved this by
concatenating a naive de Sitter transformation with the compensating
gauge transformation needed to restore the gauge condition \cite{Kleppe}.

\subsection{Tensor Basis}

Because $y(x;x')$ is de Sitter invariant, so too are covariant
derivatives of it. With the metrics $g_{\mu\nu}(x)$ and
$g_{\mu\nu}(x')$, the first three derivatives of $y(x;x')$ furnish a
convenient basis of de Sitter invariant bi-tensors \cite{KW1},
\begin{eqnarray}\label{yb1}
\frac{\partial y(x;x')}{\partial x^{\mu}} & = & H a \Bigl(y
\delta^0_{\mu}
\!+\! 2 a' H \Delta x_{\mu} \Bigr) \; , \label{dydx} \\
\frac{\partial y(x;x')}{\partial {x'}^{\nu}} & = & H a' \Bigl(y
\delta^0_{\nu}
\!-\! 2 a H \Delta x_{\nu} \Bigr) \; , \label{dydz} \\
\label{yb3}
\frac{\partial^2 y(x;x')}{\partial x^{\mu} \partial {x'}^{\nu}} & = &
H^2 a a' \Bigl(y \delta^0_{\mu} \delta^0_{\nu} \!+\! 2 a' H
\Delta x_{\mu} \delta^0_{\nu} \!-\! 2 a \delta^0_{\mu} H \Delta
x_{\nu} \!-\! 2 \eta_{\mu\nu}\Bigr) \; . \qquad \label{dydxdx'}
\end{eqnarray}
Here and subsequently we define $\Delta x_{\mu} \equiv \eta_{\mu\nu}
(x \!-\!x')^{\nu}$. Acting covariant derivatives generates more basis
tensors, for example \cite{KW1},
\begin{equation}
\frac{D^2 y(x;x')}{Dx^{\mu} Dx^{\nu}}
= H^2 (2 \!-\!y) g_{\mu\nu}(x) \quad , \quad \label{covdiv}
\frac{D^2 y(x;x')}{D {x'}^{\mu} D {x'}^{\nu}} = H^2 (2 \!-\!y)
g_{\mu\nu}(x') \; .
\end{equation}
The contraction of any pair of the basis tensors also produces more
basis tensors \cite{KW1},
\begin{eqnarray}
g^{\mu\nu}(x) \frac{\partial y}{\partial x^{\mu}} \frac{\partial
y}{\partial x^ {\nu}} & = & H^2 \Bigl(4 y - y^2\Bigr) =
g^{\mu\nu}(x') \frac{\partial y}{
\partial {x'}^{\mu}} \frac{\partial y}{\partial {x'}^{\nu}} \; ,
\label{contraction1}\\
g^{\mu\nu}(x) \frac{\partial y}{\partial x^{\nu}} \frac{\partial^2 y}{
\partial x^{\mu} \partial {x'}^{\sigma}} & = & H^2 (2-y) \frac{\partial y}{
\partial {x'}^{\sigma}} \; ,
\label{contraction2}\\
g^{\rho\sigma}(x') \frac{\partial y}{\partial {x'}^{\sigma}}
\frac{\partial^2 y}{\partial x^{\mu} \partial {x'}^{\rho}} & = & H^2
(2-y) \frac{\partial y}{\partial x^{\mu}} \; ,
\label{contraction3}\\
g^{\mu\nu}(x) \frac{\partial^2 y}{\partial x^{\mu} \partial
{x'}^{\rho}} \frac{\partial^2 y}{\partial x^{\nu} \partial {x'}^{\sigma}}
& = & 4 H^4 g_{\rho\sigma}(x') - H^2 \frac{\partial y}{\partial
{x'}^{\rho}} \frac{\partial y}{\partial {x'}^{\sigma}} \; ,
\label{contraction4}\\
g^{\rho\sigma}(x') \frac{\partial^2 y}{\partial x^{\mu}\partial
{x'}^{\rho}} \frac{\partial^2 y}{\partial x^{\nu} \partial {x'}^{\sigma}}
& = & 4 H^4 g_{\mu\nu}(x) - H^2 \frac{\partial y}{\partial x^{\mu}}
\frac{\partial y}{\partial x^{\nu}} \; . \label{contraction5}
\end{eqnarray}

The tensor structure of de Sitter breaking terms requires
derivatives of the quantity $u(x;x') \equiv \ln(a a')$,
\begin{equation}
\frac{\partial u}{\partial x^{\mu}} = H a \delta^0_{\mu} \qquad ,
\qquad \frac{\partial u}{\partial {x'}^{\rho}} = H a' \delta^0_{\rho}
\; . \label{uders}
\end{equation}
Covariant derivatives of the new tensors involve some extra
identities in addition to those of $y(x;x')$ \cite{MTW3},
\begin{equation}
\frac{D^2 u}{D x^{\mu} D x^{\nu}} = -H^2 g_{\mu\nu}(x) -
\frac{\partial u}{\partial x^{\mu}} \frac{\partial u}{\partial
x^{\nu}} \; , \; \frac{D^2 u}{D {x'}^{\mu} D {x'}^{\nu}} = -H^2
g_{\mu\nu}(x') - \frac{\partial u}{\partial {x'}^{\mu}} \frac{\partial
u}{\partial {x'}^{\nu}} \; .
\end{equation}
There are also some new contraction identities,
\begin{eqnarray}
g^{\mu\nu}(x) \frac{\partial u}{\partial x^{\mu}} \frac{\partial
u}{\partial x^{\nu}} & = & - H^2 = g^{\rho\sigma}(x') \frac{\partial
u}{\partial {x'}^{\rho}} \frac{\partial u}{\partial {x'}^{\sigma}} \; , \\
g^{\mu\nu}(x) \frac{\partial u}{\partial x^{\mu}} \frac{\partial
y}{\partial x^{\nu}} & = & - H^2 \Bigl[ y \!-\! 2
+ 2 \frac{a'}{a} \Bigr] \; , \\
g^{\rho\sigma}(x') \frac{\partial u}{\partial {x'}^{\rho}}
\frac{\partial y}{\partial {x'}^{\sigma}} & = & - H^2 \Bigl[ y \!-\! 2
+ 2 \frac{a}{a'} \Bigr] \; , \\
g^{\mu\nu}(x) \frac{\partial u}{\partial x^{\mu}} \frac{\partial^2
y}{\partial x^{\nu} \partial {x'}^{\rho}} & = & - H^2 \Bigl[
\frac{\partial y}{\partial {x'}^{\rho}} + 2 \frac{a'}{a}
\frac{\partial u}{\partial {x'}^{\rho}} \Bigr] \; , \\
g^{\rho\sigma}(x') \frac{\partial u}{\partial {x'}^{\rho}}
\frac{\partial^2 y}{\partial x^{\mu} \partial {x'}^{\sigma}} & = & -
H^2 \Bigl[ \frac{\partial y}{\partial x^{\mu}} + 2 \frac{a}{a'}
\frac{\partial u}{\partial x^{\mu}} \Bigr] \; .
\end{eqnarray}

Finally, we should explain the relation of our tensor basis to the
one employed by mathematical physicists. Their literature obviously
includes no mention of the de Sitter breaking tensors $\partial
u/\partial x^{\mu}$ and $\partial u/\partial {x'}^{\mu}$, however,
there are also significant differences in the de Sitter invariant
sector. Our motivation for employing derivatives of the length
function $y(x;x')$ is to simplify loop computations which involve
derivatives of propagators. That is not a significant consideration
for mathematical physicists because their literature is devoid of
such computations; the only quantum gravitational loop computations
so far made on de Sitter background \cite{TW3,TW4,TW5,MW1,KW2} use
our propagator. The issue of greater importance to mathematical
physicists is the geometrical significance of the tensor basis. In
place of $\partial y/\partial x^{\mu}$ and $\partial y/\partial
{x'}^{\mu}$, they accordingly employ derivatives of the geodetic
length function $\ell(x;x')$ (which is known as ``$\mu$'' in their
literature),
\begin{eqnarray}
n_{\mu} & \equiv & \frac{\partial \ell(x;x')}{\partial x^{\mu}} =
\frac{\frac{\partial y}{\partial x^{\mu}} }{H \sqrt{4 y \!-\! y^2}} \; , 
\label{nun} \\
n_{\mu'} & \equiv & \frac{\partial \ell(x;x')}{\partial {x'}^{\mu'}} =
\frac{\frac{\partial y}{\partial {x'}^{\mu'}} }{H \sqrt{4 y \!-\! y^2}} \; .
\label{npr}
\end{eqnarray}
(Note the mathematical physics notation in which unprimed indices
belong to the tangent space at $x^{\mu}$ and primed indices belong
to the ${x'}^{\mu}$ tangent space.) In place of the mixed second
derivative $\partial^2y/\partial x^{\mu} \partial {x'}^{\nu}$,
mathematical physicists prefer the parallel transport matrix,
\begin{equation}
g_{\mu \nu'} = -\frac1{2 H^2} \Biggl[ \frac{\partial^2 y}{\partial x^{\mu}
\partial {x'}^{\nu'}} + \frac1{4 \!-\! y} \frac{\partial y}{\partial x^{\mu}}
\frac{\partial y}{\partial {x'}^{\nu'}} \Biggr] \; . \label{par}
\end{equation}

\section{Doing the Math}

The purpose of this section is to perform the actual computation. We
begin by exploiting conformal invariance to write the Weyl-Weyl
correlator as a series of permutations and traces of four ordinary
derivatives of the graviton propagator. We then express the index
factors of the graviton propagator using the tensor basis of the
previous section. The next step is to reduce the four ordinary
derivatives of the various scalar propagator functions to a standard
form based on the same tensor basis. The final step is to note that
the standard permutations and traces remove all the noncovariant
tensors, leaving only a linear combination of three de Sitter
invariant tensors times exceptionally simple scalar factors. We also
compare with the result of Kouris \cite{Kouris}, and we take the
coincidence limit using dimensional regularization.

\subsection{Exploiting Conformal Invariance}

Recall the relation (\ref{hdef}) between the conformally transformed
metric $\widetilde{g}_{\mu\nu}$ and the full metric $g_{\mu\nu} = a^2
\widetilde{g}_{\mu\nu}$. Let $C_{\alpha\beta\gamma\delta}$ and
$\widetilde{C}_{\alpha\beta\gamma\delta}$ stand for the Weyl tensors
constructed from each metric, with their indices raised and lowered
by the appropriate metric. Because the Weyl tensor is conformally
invariant with one index raised we have,
\begin{equation}
{C^\alpha}_{\beta\rho\sigma} = \widetilde{C}^\alpha_{~\beta\rho\sigma}
\qquad \Longrightarrow \qquad
C_{\alpha\beta\rho\sigma} =a^2 \widetilde{C}_{\alpha\beta\rho\sigma} \; .
\label{weylNonInv}
\end{equation}
As a consequence the correlation function of two Weyl tensors takes the form,
\begin{equation}
\label{WWC1}
\Bigl\langle \Omega \Bigl\vert C_{\alpha\beta\gamma\delta}(x)
C_{\mu\nu\rho\sigma}(x') \Bigr\vert \Omega \Bigr\rangle = a^2 a'^2
\Bigl\langle \Omega \Bigl\vert \widetilde{C}_{\alpha\beta\gamma\delta}(x)
\widetilde{C}_{\mu\nu\rho\sigma}(x') \Bigr\vert \Omega \Bigr\rangle \; .
\end{equation}

The advantage of conformal invariance becomes apparent when we express
the Weyl tensor in terms of the Riemann tensor ($R^{\rho}_{~\sigma\mu\nu}
\equiv \partial_{\mu} \Gamma^{\rho}_{~\nu\sigma} + \Gamma^{\rho}_{~\mu\alpha}
\Gamma^{\alpha}_{~\nu\sigma} - \mu \leftrightarrow \nu$) and its traces
$R_{\mu\nu} \equiv R^{\rho}_{~ \mu\rho\nu}$ and $R \equiv g^{\mu\nu}
R_{\mu\nu}$,
\begin{eqnarray}
\label{weylDef}
\lefteqn{ {C}_{\alpha\beta\gamma\delta} =
{R}_{\alpha\beta\gamma\delta} - \frac1{D \!-\! 2} \Bigl(
{g}_{\alpha\gamma} R_{\beta\delta} \!-\!
{g}_{\gamma\beta} R_{\delta\alpha} \!+\!
{g}_{\beta\delta} R_{\alpha\gamma} \!-\!
{g}_{\delta\alpha} R_{\gamma\beta}\Bigr) } \nonumber \\
& & \hspace{5.5cm} + \frac1{(D\!-\!2) (D\!-\!1)} \Bigl(
g_{\alpha\gamma} g_{\beta\delta} \!-\!
g_{\alpha\delta} g_{\beta\gamma} \Bigr) R \; . \qquad
\end{eqnarray}
Of course the same relation (\ref{weylDef}) gives the conformally
transformed Weyl tensor in terms of the conformally transformed metrics
and curvatures. But whereas the de Sitter background of $g_{\mu\nu}$
is curved, the background value of the conformally transformed metric
is flat $\widetilde{g}_{\mu\nu} = \eta_{\mu\nu} + \kappa h_{\mu\nu}$.
This makes it very simple to extract the linearized piece,
\begin{equation}
\label{LinRiem}
\widetilde{R}_{\alpha\beta\gamma\delta}(x) = -\frac{\kappa}{2}
\Bigl( h_{\beta\delta , \alpha\gamma} - h_{\delta\alpha , \gamma\beta}
+ h_{\alpha\gamma , \beta\delta} - h_{\gamma\beta , \delta\alpha} \Bigr) +
\mathcal{O}(\kappa^2) \; .
\end{equation}

It remains to describe the index algebra needed to convert the quadruply
differentiated propagator into the linearized Weyl-Weyl correlator
\begin{equation}
\frac{\kappa^2}{4} \partial_{\alpha} \partial_{\gamma} \partial_{\mu}'
\partial_{\rho}' i \Bigl[ \mbox{}_{\beta\delta} \Delta_{\nu\sigma}
\Bigr](x;x') \longrightarrow \Bigl\langle \Omega \Bigl\vert
\widetilde{C}_{\alpha\beta\gamma\delta}(x)
\widetilde{C}_{\mu\nu\rho\sigma}(x') \Bigr\vert \Omega \Bigr\rangle
+ O(\kappa^4) \; . \label{transition}
\end{equation}
We distinguish two steps:
\begin{itemize}
\item{{\it Riemannization}, in which the linearized (and conformally
transformed) Riemann-Riemann correlator is formed; and }
\item{{\it Weylization}, in which the traces are subtracted to give the
linearized Weyl-Weyl correlator.}
\end{itemize}
It is useful to define Riemannization generally for any 8-index bi-tensor
``seed'' with the same algebraic symmetries as the quadruply differentiated
propagator on the left hand side of (\ref{transition}). From expression
(\ref{LinRiem}) we infer,
\begin{equation}
\label{Rdef}
{\rm Riem}\Bigl[ ({\rm seed})_{\alpha\beta\gamma\delta
\mu\nu\rho\sigma} \Bigr] \equiv
\mathcal{R}_{\alpha\beta\gamma\delta}^{~~~~~ \epsilon\zeta\kappa\lambda}
\times \mathcal{R}_{\mu\nu\rho\sigma}^{~~~~~ \theta\phi\psi\omega} \times
({\rm seed})_{\epsilon\zeta\kappa\lambda\theta\phi\psi\omega} \; ,
\end{equation}
where,
\begin{equation}
\mathcal{R}_{\alpha\beta\gamma\delta}^{~~~~~\epsilon\zeta\kappa\lambda}
\equiv \delta^{\epsilon}_{\alpha} \delta^{\kappa}_{\gamma}
\delta^{\zeta}_{\beta} \delta^{\lambda}_{\delta}
- \delta^{\epsilon}_{\gamma} \delta^{\kappa}_{\beta}
\delta^{\zeta}_{\delta} \delta^{\lambda}_{\alpha}
+ \delta^{\epsilon}_{\beta} \delta^{\kappa}_{\delta}
\delta^{\zeta}_{\alpha} \delta^{\lambda}_{\gamma}
- \delta^{\epsilon}_{\delta} \delta^{\kappa}_{\alpha}
\delta^{\zeta}_{\gamma} \delta^{\lambda}_{\beta} \; .
\end{equation}
Weylization can be defined similarly on any 8-index bi-tensor seed
with the algebraic symmetries of the product of two Riemann tensors,
\begin{equation}
{\rm Weyl}\Bigl[ ({\rm seed})_{\alpha\beta\gamma\delta
\mu\nu\rho\sigma} \Bigr] \equiv
\mathcal{C}_{\alpha\beta\gamma\delta}^{~~~~~ \epsilon\zeta\kappa\lambda}
\times \mathcal{C}_{\mu\nu\rho\sigma}^{~~~~~ \theta\phi\psi\omega} \times
({\rm seed})_{\epsilon\zeta\kappa\lambda\theta\phi\psi\omega} \; .
\label{Weylization}
\end{equation}
From expression (\ref{weylDef}) we infer,
\begin{eqnarray}
\label{cDef}
\lefteqn{ \mathcal{C}_{\alpha\beta\gamma\delta}^{~~~~~ \epsilon
\zeta\kappa\lambda} \equiv \delta^{\epsilon}_{\alpha} \delta^{\zeta}_{\beta}
\delta^{\kappa}_{\gamma} \delta^{\lambda}_{\delta} -
\Bigl[\eta_{\alpha\gamma} \delta^{\zeta}_{\beta} \delta^{\lambda}_{\delta}
- \eta_{\gamma\beta} \delta^{\zeta}_{\delta} \delta^{\lambda}_{\alpha}
+ \eta_{\beta\delta} \delta^{\zeta}_{\alpha} \delta^{\lambda}_{\gamma}
- \eta_{\delta\alpha} \delta^{\zeta}_{\gamma} \delta^{\lambda}_{\beta}
\Bigr] \frac{ \eta^{\epsilon\kappa}}{D \!-\! 2} } \nonumber \\
& & \hspace{6cm} + \Bigl[ \eta_{\alpha\gamma} \eta_{\beta\delta} \!-\!
\eta_{\alpha\gamma} \eta_{\beta\delta}\Bigr] \frac{\eta^{\epsilon\kappa}
\eta^{\zeta\lambda}}{(D \!-\! 2) (D \!-\!1) } \; . \qquad
\end{eqnarray}

The operations of Riemannization and Weylization give a simple form
for the linearized Weyl-Weyl correlator,
\begin{eqnarray}
\label{WWC3}
\lefteqn{\Bigl\langle \Omega \Bigl\vert C_{\alpha\beta\gamma\delta}(x)
C_{\mu\nu\rho\sigma}(x') \Bigr\vert \Omega \Bigr\rangle } \nonumber \\
& & \hspace{1cm} = \frac{\kappa^2}4 \, a^2 {a'}^2 \, {\rm Weyl}\Biggr(
{\rm Riem}\Biggr[ \partial_{\alpha} \partial_{\gamma} \partial'_{\mu}
\partial'_{\rho} \, i\Bigr[\mbox{}_{\beta\delta} \Delta_{\nu\sigma}
\Bigr](x;x') \Biggr] \Biggr) + O(\kappa^4) \; . \qquad
\end{eqnarray}
From expression (\ref{gprop}) for the graviton propagator, and the
fact that the index factors $[\mbox{}_{\beta\delta} T^I_{\nu\sigma}]$
are constant in our gauge, we can write,
\begin{equation}
a^2 {a'}^2 \partial_{\alpha} \partial_{\gamma} \partial'_{\mu}
\partial'_{\rho} \, i\Bigr[\mbox{}_{\beta\delta} \Delta_{\nu\sigma}
\Bigr](x;x') = \sum_{I = A,B,C} a^2 {a'}^2 \Bigl[ \mbox{}_{\beta\delta}
T^I_{\nu\sigma} \Bigr] \times \partial_{\alpha} \partial_{\gamma}
\partial_{\mu}' \partial_{\rho}' i\Delta_I(x;x') \; . \label{seed1}
\end{equation}
In the next two subsections we will derive expressions for first
$a^2 {a'}^2 [\mbox{}_{\beta\delta} T^I_{\nu\sigma}]$ and then
$\partial_{\alpha} \partial_{\gamma} \partial_{\mu}' \partial_{\rho}'
i\Delta_I(x;x')$.

Several comments are in order before we close this subsection. First,
Riemannization is the ``standard permutation'' defined decades ago in
a study of invariant Green's functions \cite{TW7}. A result from that
work which will facilitate subsequent analysis is that the Riemannization
of any seed which is symmetric on the index pairs $(\alpha,\gamma)$,
$(\beta,\delta)$, $(\mu,\rho)$ and $(\nu,\sigma)$ will possess all the
algebraic symmetries of a Riemann tensor at each point,
\begin{equation}
R_{\alpha\beta\gamma\delta} = -R_{\beta\alpha\gamma\delta} =
-R_{\alpha\beta\delta\gamma} = R_{\gamma\delta\alpha\beta} =
R_{\alpha\gamma\beta\delta} - R_{\alpha\delta\beta\gamma} \; .
\end{equation}
The second point is that the Weyl tensor possesses the additional
algebraic symmetry of being traceless on any two indices, and the
additional differential symmetry of being transverse,
\begin{equation}
D^{\alpha} C_{\alpha\beta\gamma\delta} = 0 \; . \label{transverse}
\end{equation}
Of course it is the full covariant derivative operator that appears
in (\ref{transverse}), but the covariant derivative of the de Sitter
background must annihilate the linearized Weyl-Weyl correlator.
Third, every factor of the Minkowski metric in (\ref{cDef}) is
accompanied by an inverse metric, so we could just have easily
expressed this tensor in terms of the de Sitter background metric,
\begin{equation}
\eta_{\alpha\gamma} \eta^{\epsilon\kappa} = a^2 \eta_{\alpha\gamma}
\times \frac1{a^2} \eta^{\epsilon\kappa} \equiv g_{\alpha\gamma}(x)
\times g^{\epsilon\kappa}(x) \; .
\end{equation}
Our final point is evident in the last equation: because we no longer
need the full metric $g_{\mu\nu} = a^2 (\eta_{\mu\nu} + \kappa h_{\mu\nu})$,
we will henceforth employ the symbol ``$g_{\mu\nu}$'' to denote the de
Sitter background metric, $g_{\mu\nu} \equiv a^2 \eta_{\mu\nu}$.

\subsection{Standard Form for Tensor Structures}

The de Sitter invariant part of the index factors can be written in
terms of the $y$-basis introduced in Section 2.3. To keep the tensor
factors dimensionless we employ the notation,
\begin{eqnarray}
\label{defs}
\mathcal{Y}_{\mu} \equiv \frac1{H} \frac{\partial y}{\partial x^{\mu}}
\qquad , \qquad \mathcal{Y}'_{\nu} \equiv \frac1{H}
\frac{\partial y}{\partial {x'}^{\nu}} \\
\label{defs2} \mathcal{R}_{\mu\nu}(x;x') \equiv -\frac{1}{2H^2}
\frac{\partial^2y}{\partial x^\mu\partial x'^\nu} \; .
\end{eqnarray}
The analogous dimensionless de Sitter breaking tensors are,
\begin{equation}
\label{dbT}
\mathcal{T}_{\mu} \equiv \frac1{H} \frac{\partial u}{\partial x^{\mu}}
= a \delta^0_\mu \qquad , \qquad
\mathcal{T}'_{\nu} \equiv \frac1{H} \frac{\partial u}{\partial {x'}^{\nu}}
= a' \delta^0_\nu \; . \label{taus}
\end{equation}

The key to extracting the invariant parts of the various index factors
(\ref{TA}-\ref{TC}) is to note that they involve purely temporal tensors
such as $\delta^0_{\mu}$ and purely spatial tensors such as
$\overline{\eta}_{\mu\nu} \equiv \eta_{\mu\nu} + \delta^0_{\mu}
\delta^0_{\nu}$. Of course the temporal factors can be represented
using (\ref{taus}). The purely spatial metric can either involve two
indices from the same point, or from both points. If the indices are
from the same point we can represent it using the purely spatial tangent
matrix introduced in \cite{KMW},
\begin{equation}
g^{\perp}_{\beta\delta}(x) \equiv g_{\beta\delta}(x) + \mathcal{T}_{\beta}
\mathcal{T}_{\delta} = a^2 \overline{\eta}_{\beta\delta} \quad , \quad
g^{\perp}_{\nu\sigma}(x') \equiv g_{\nu\sigma}(x') + \mathcal{T}'_{\nu}
\mathcal{T}'_{\sigma} = {a'}^2 \overline{\eta}_{\nu\sigma} \; .
\end{equation}
The case of mixed indices is given by \cite{KMW},
\begin{eqnarray}
\mathcal{R}^{\perp}_{\mu\nu}(x;x') \equiv\mathcal{R}_{\mu\nu}(x;x') \!+\!
\frac12 \mathcal{Y}_\mu\mathcal{T}'_\nu \!+\! \frac12 \mathcal{T}_{\mu}
\mathcal{Y}'_{\nu} \!+\! \frac{(2-y)}{2} \mathcal{T}_{\mu}
\mathcal{T}'_{\nu} = a a' \overline{\eta}_{\mu\nu}
\end{eqnarray}
With these definitions the three tensor factors take the form,
\begin{eqnarray}
\label{tfSt}
a^2 {a'}^2 \Bigl[ \mbox{}_{\beta\delta} T^A_{\nu\sigma}\Bigr] & \!\!\!\!=
\!\!\!\! & \mathcal{R}^{\perp}_{\beta\nu} \mathcal{R}^{\perp}_{\delta\sigma} +
\mathcal{R}^{\perp}_{\beta\sigma} \mathcal{R}^{\perp}_{\delta\nu} -
\frac2{D \!-\! 3} \, g^{\perp}_{\beta\delta}(x) g^{\perp}_{\nu\sigma}(x') \\
a^2 {a'}^2 \Bigl[ \mbox{}_{\beta\delta} T^B_{\nu\sigma}\Bigr] & \!\!\!\!=
\!\!\!\! & -4 \mathcal{T}_{(\beta} \mathcal{R}^{\perp}_{\delta) (\nu}
\mathcal{T}'_{\sigma)} \\
a^2 {a'}^2 \Bigl[ \mbox{}_{\beta\delta} T^C_{\nu\sigma}\Bigr] & \!\!\!\! =
\!\!\!\! & \frac2{(D \!-\! 3) (D \!-\! 2)} \Bigl[ (D \!-\! 2)
\mathcal{T}_{\beta} \mathcal{T}_{\delta} + g_{\beta\delta}\Bigr]
\Bigl[ (D \!-\!2) \mathcal{T}'_{\nu} \mathcal{T}'_{\sigma} +
g_{\nu\sigma}' \Bigr] . \quad
\end{eqnarray}

\subsection{Standard Form for Derivatives}

We can perform a similar reduction for the factor $\partial_{\alpha}
\partial_{\gamma} \partial'_{\mu} \partial'_{\rho}\, i\Delta_I(x;x')$
in (\ref{seed1}). The $B$-type and $C$-type propagators are de Sitter
invariant functions of $y(x;x')$, and taking two mixed derivatives of
the $A$-type propagator eliminates its de Sitter breaking term.
Thus, after acting these first two derivatives we can write,
\begin{equation}
\partial_{\alpha} \partial_{\gamma} \partial'_{\mu} \partial'_{\rho}
\, i\Delta_I(x;x') = \partial_{\alpha} \partial'_{\mu} \Biggl\{ I'(y)
\frac{\partial^2 y}{\partial x^{\gamma} \partial {x'}^{\rho}} +
I''(y) \frac{\partial y}{\partial x^{\gamma}} \frac{\partial y}{\partial
{x'}^{\rho}} \Biggr\} \; .
\end{equation}
Acting the remaining two derivatives produces noninvariant terms,
\begin{eqnarray}
\label{4ds}
\lefteqn{\partial_{\alpha} \partial_{\gamma} \partial'_{\mu}
\partial'_{\rho} i\Delta_I(x;x') = \Biggl[ \frac{\partial^4y}{\partial
x^{\alpha} \partial x^{\gamma} \partial x'^{\mu} \partial x'^{\rho}}
\Biggr] I'(y) + \Biggl[ \frac{\partial^2y}{\partial x^{\alpha}
\partial x^{\gamma}} \frac{\partial^2y}{\partial x'^{\mu} \partial x'^{\rho}}}
\Biggr. \nonumber \\
& & \hspace{0cm} + \Biggl. 4 \frac{\partial^2y}{\partial x^{\alpha}
\partial x'^{(\mu}} \frac{\partial^2 y}{\partial x'^{\rho)}
\partial x^{\gamma}} \frac{\partial y}{\partial x^{(\alpha}}
\frac{\partial^3 y}{\partial x^{\gamma)} \partial x'^{\mu} \partial x'^{\rho}}
+ 2 \frac{\partial^3y}{\partial x^{\alpha} \partial x^{\gamma}
\partial x'^{(\mu}} \frac{\partial y}{\partial x'^{\rho)}} \Biggr]
I''(y) \nonumber \\
& & \hspace{0cm} + \Biggr[ 4 \frac{\partial y}{\partial x^{(\alpha}}
\frac{\partial^2 y}{\partial x^{\gamma)} \partial x'^{(\mu}}
\frac{\partial y}{\partial x'^{\rho)}} + \frac{\partial y}{\partial x^{\alpha}}
\frac{\partial y}{\partial x^{\gamma}} \frac{\partial^2y}{\partial x'^{\mu}
\partial x'^{\rho}} + \frac{\partial^2y}{\partial x^{\alpha}
\partial x^{\gamma}} \frac{\partial y}{\partial x'^{\mu}}
\frac{\partial y}{\partial x'^{\rho}} \Biggl] I'''(y) \nonumber \\
& & \hspace{5cm} + \Biggl[\frac{\partial y}{\partial x^{\alpha}}
\frac{\partial y}{\partial x^{\gamma}} \frac{\partial y}{\partial x'^{\mu}}
\frac{\partial y}{\partial x'^{\rho}} \Biggr] I''''(y) \; .
\end{eqnarray}
All noninvariance comes from acting two derivatives at the same spacetime
point. We can express these derivatives in standard form,
\begin{eqnarray}
\label{mult}
\frac{\partial^2 y}{\partial x^{\alpha} \partial x^{\gamma}} & = &
2 H^2 \Biggl\{ \frac{a'}{a} g_{\alpha\gamma}(x) + \mathcal{T}_{(\alpha}
\mathcal{Y}_{\gamma)} \Biggr\} \; , \nonumber\\
\frac{\partial^2 y}{\partial x'^{\mu} \partial x'^{\rho}} & = &
2 H^2 \Biggl\{ \frac{a}{a'} g_{\mu\rho}(x') + \mathcal{T}'_{(\mu}
\mathcal{Y}'_{\rho)} \Biggr\} \; , \nonumber\\
\frac{\partial^3 y}{\partial x^{\alpha} \partial x^{\gamma}
\partial {x'}^{\mu}} & = & 2 H^3 \Biggl\{ \frac{a'}{a} g_{\alpha\gamma}(x)
\mathcal{T}'_{\mu} - 2 \mathcal{T}_{(\alpha} \mathcal{R}_{\gamma) \mu}
\Biggr\} \; , \nonumber\\
\frac{\partial^3 y}{\partial x^{\alpha} \partial x'^{\mu}
\partial {x'}^{\rho}} & = & 2 H^3 \Biggl\{ \frac{a}{a'} g_{\mu\rho}(x')
\mathcal{T}_{\alpha} - 2  \mathcal{R}_{\alpha(\mu} \mathcal{T}'_{\rho)}
\Biggr\} \; , \nonumber\\
\frac{\partial^4 y}{\partial x^{\alpha} \partial x^{\gamma}
\partial {x'}^{\mu} \partial {x'}^{\rho}} & = & 4 H^4 \Biggl\{
\frac{a}{a'} \mathcal{T}_{\alpha} \mathcal{T}_{\gamma} g_{\mu\rho}(x')
\nonumber \\
& & \hspace{2.5cm} +
\frac{a'}{a} g_{\alpha\gamma}(x) \mathcal{T}'_{\mu} \mathcal{T}'_{\rho}
- 2 \mathcal{T}_{(\alpha} \mathcal{R}_{\gamma) (\mu} \mathcal{T}'_{\rho)}
\Biggr\} \; . \qquad
\end{eqnarray}

\subsection{The Final Result}

Most of the subsequent analysis was made using the symbolic manipulation 
program {\it Mathematica}, but it is of course advantageous to simplify 
even computer calculations to make them run more efficiently and
transparently. It is evident that Riemannizing and then Weylizing
our original seed (\ref{seed1}) will produce a huge number of terms,
many of which are permutations and traces of the same seed tensor
times some function of $y$. Rather than process this unwieldy form
all the way through Weylization, we expressed the Riemannized result
as a linear combination of the rather small number of tensors which
possess the algebraic symmetries of the product of two Riemann
tensors. It turns out there are only nine independent invariant
tensors with these symmetries \cite{TW7}. There are many more
noninvariant tensors, but very few of these actually occur.

A further simplification is to break the Riemannized result into
those terms ${\rm R}^{\rm
(g)}_{\alpha\beta\gamma\delta\mu\nu\rho\sigma}$ which contain one or
more factors of the de Sitter metric and those ${\rm R}^{\rm
(ng)}_{\alpha\beta\gamma\delta\mu\nu\rho\sigma}$ which do not,
\begin{equation}
\frac{\kappa^2}{4} a^2 {a'}^2 {\rm Riem}\Biggl[ \partial_{\alpha}
\partial_{\gamma} \partial_{\mu}' \partial_{\rho}' i\Bigl[
\mbox{}_{\beta\delta} \Delta_{\nu\sigma}\Bigr](x;x') \Biggr] = {\rm
R}^{\rm (ng)}_{\alpha\beta\gamma\delta\mu\nu\rho\sigma} + {\rm
R}^{\rm (g)}_{\alpha\beta\gamma\delta\mu\nu\rho\sigma} \; .
\end{equation}
This is useful because Weylization can only change the metric terms.
Of course there is the additional advantage that the number of
independent tensors needed to represent the nonmetric terms is
smaller. Without the metric there are only three invariant tensors
with the algebraic symmetries of a product of two Riemann tensors
\cite{TW7}. We can represent them by Riemannizing the seeds,
\begin{eqnarray}
\sigma^{(1)}_{\alpha\beta\gamma\delta\mu\nu\rho\sigma} & = & 2
\frac{\partial^2 y}{\partial x^{\alpha} \partial x'^{(\mu}}
\frac{\partial^2 y}{\partial {x'}^{\rho)} \partial x^{\gamma}} \times
\frac{\partial^2 y}{\partial x^{\beta} \partial x'^{(\nu}}
\frac{\partial^2 y}{\partial {x'}^{\sigma)} \partial x^{\delta}} \; ,
\label{cov1} \\
\sigma^{(2)}_{\alpha\beta\gamma\delta\mu\nu\rho\sigma} & = & 8
\frac{\partial y}{\partial x^{(\alpha}}
\frac{\partial^2 y}{\partial x^{\gamma)} \partial x'^{(\mu}}
\frac{\partial y}{\partial x'^{\rho)}} \times
\frac{\partial^2 y}{\partial x^{\beta} \partial x'^{(\nu}}
\frac{\partial^2 y}{\partial {x'}^{\sigma)}\partial x^{\delta}} \; ,
\label{cov2} \\
\sigma^{(3)}_{\alpha\beta\gamma\delta\mu\nu\rho\sigma} & = & 2
\frac{\partial y}{\partial x^{\alpha}} \frac{\partial y}{\partial x^{\gamma}}
\frac{\partial y}{\partial x'^{\mu}} \frac{\partial y}{\partial x'^{\rho}}
\times \frac{\partial^2 y}{\partial x^{\beta} \partial x'^{(\nu}}
\frac{\partial^2 y}{\partial {x'}^{\sigma)} \partial x^{\delta}} \; .
\label{cov3}
\end{eqnarray}
Although many de Sitter breaking, nonmetric tensors are conceivable, it
turns out that only three occur. They derive from Riemannizing the
seeds,
\begin{eqnarray}
\sigma^{(4)}_{\alpha\beta\gamma\delta\mu\nu\rho\sigma} & = & 2
\frac{\partial u}{\partial x^{\alpha}} \frac{\partial u}{\partial x^{\gamma}}
\frac{\partial u}{\partial x'^{\mu}} \frac{\partial u}{\partial x'^{\rho}}
\times \frac{\partial^2 y}{\partial x^{\beta} \partial x'^{(\nu}}
\frac{\partial^2 y}{\partial {x'}^{\sigma)} \partial x^{\delta}} \; ,
\label{noncov1} \\
\sigma^{(5)}_{\alpha\beta\gamma\delta\mu\nu\rho\sigma} & = & 4
\frac{\partial u}{\partial x^{\alpha}} \frac{\partial u}{\partial
x^{\gamma}} \frac{\partial u}{\partial x'^{\mu}} \frac{\partial
u}{\partial x'^{\rho}} \times \frac{\partial y}{\partial x^{(\beta}}
\frac{\partial^2 y}{\partial x^{\delta)} \partial x'^{(\nu}}
\frac{\partial y}{\partial x'^{\sigma)}} \; , \label{noncov2} \\
\sigma^{(6)}_{\alpha\beta\gamma\delta\mu\nu\rho\sigma} & = &
\frac{\partial y}{\partial x^{\alpha}} \frac{\partial y}{\partial x^{\gamma}}
\frac{\partial y}{\partial x'^{\mu}} \frac{\partial y}{\partial x'^{\rho}}
\times \frac{\partial u}{\partial x^{\beta}}
\frac{\partial u}{\partial x^{\delta}} \frac{\partial u}{\partial x'^{\nu}}
\frac{\partial u}{\partial x'^{\sigma}} \; . \label{noncov3}
\end{eqnarray}

We extracted the corresponding coefficients of seeds $\sigma^{(1)}$,
$\sigma^{(2)}$ and $\sigma^{(3)}$ in ${\rm R}^{\rm
(ng)}_{\alpha\beta\gamma\delta\mu\nu\rho\sigma}$ and they have the
wonderfully simple forms,
\begin{equation}
\label{3coef} c_1 = \frac{\kappa^2}{8} A''(y) \;\;\; , \;\;\; c_2 =
\frac{\kappa^2}{16} A'''(y) \;\;\; , \;\;\; c_3 =
\frac{\kappa^2}{16} A''''(y) \; .
\end{equation}
Even better are the results we obtained for the coefficients of the
noninvariant tensors (\ref{noncov1}-\ref{noncov3}),
\begin{eqnarray}
\label{vanCoef} c_4 & \!\!\!\!=\!\!\!\! & -\frac{\kappa^2}{8 (D
\!-\! 3)} \Bigl[ (4y \!-\! y^2) A''(y) \!+\! D (2 \!-\!y) A'(y)
\!-\! (D \!-\! 1) k \Bigr] \; , \\
c_5 & \!\!\!\!=\!\!\!\! & -\frac{\kappa^2}{8(D \!-\! 3)} \Bigl[ (4y
\!-\! y^2) A'''(y) \!+\! (D \!+\! 2) (2 \!-\! y) A''(y) \!-\! D
A'(y) \Bigr] \; , \\
c_6 & \!\!\!\!=\!\!\!\! & -\frac{\kappa^2}{8 (D \!-\! 3)} \Bigl[ (4y
\!-\!y^2) A''''(y) \!+\! (D \!+\! 4) (2 \!-\!y) A'''(y) \!-\! 2 (D
\!+\! 1) A''(y) \Bigr] \; . \qquad
\end{eqnarray}
Note that the coefficient $c_4$ is proportional to the differential
equation (\ref{ape}) satisfied by $A(y)$, while $c_5$ and $c_6$ are
proportional to its first and second derivatives, respectively. So
these three coefficients vanish and we can write,
\begin{equation}
\label{NOg} \rm{R}^{\rm
(ng)}_{\alpha\beta\gamma\delta\mu\nu\rho\sigma} = \sum_{k=1}^3 c_k
\times {\rm Riem}\Bigl[ \sigma^{(k)}_{\alpha\beta\gamma\delta
\mu\nu\rho\sigma} \Bigr] \; .
\end{equation}

Let us now turn to the Riemannized terms which contain one or more
factors of the de Sitter metric, ${\rm R}^{\rm
(g)}_{\alpha\beta\gamma\delta\mu\nu\rho\sigma}$. Although the list
for all possible (invariant and noninvariant) seed tensors is much
longer than the first one, it turns out that they all vanish upon
Weylization,
\begin{equation}
{\rm Weyl}\Bigl( {\rm R}^{\rm (g)}_{\alpha\beta\gamma\delta
\mu\nu\rho\sigma} \Bigr) = 0 \; .
\end{equation}
Hence the final result is just the Weylization of (\ref{NOg}).
Expressing the seed tensors (\ref{cov1}-\ref{cov3}) in our standard,
dimensionless form gives,
\begin{eqnarray}
\label{finalanswer}
\lefteqn{ \Bigl\langle \Omega \Bigl\vert C_{\alpha\beta\gamma\delta}(x) \times
C_{\mu\nu\rho\sigma}(x') \Bigr\vert \Omega \Bigr\rangle =
\kappa^2 H^4 A''(y) } \nonumber \\
& & \hspace{-.2cm} \times {\rm Weyl}\Biggl(
{\rm Riem}\Biggl[ \Bigl[ \mathcal{R}_{\alpha\mu} \mathcal{R}_{\gamma\rho}
\!+\! \mathcal{R}_{\alpha\rho} \mathcal{R}_{\gamma\mu} \Bigr] \Bigl[
\mathcal{R}_{\beta\nu} \mathcal{R}_{\delta\sigma} \!+\!
\mathcal{R}_{\beta\sigma} \mathcal{R}_{\delta\nu}\Bigr] \Biggr] \Biggr)
-2 \kappa^2 H^4 A'''(y) \nonumber \\
& & \hspace{.7cm} \times {\rm Weyl}\Biggl(
{\rm Riem}\Biggl[ \mathcal{Y}_{(\alpha} \mathcal{R}_{\gamma) (\mu}
\mathcal{Y}'_{\rho)}  \Bigl[ \mathcal{R}_{\beta\nu}
\mathcal{R}_{\delta\sigma} \!+\!  \mathcal{R}_{\beta\sigma}
\mathcal{R}_{\delta\nu}\Bigr] \Biggr] \Biggr)
+ \frac14 \kappa^2 H^4 A''''(y) \nonumber \\
& & \hspace{1.5cm} \times {\rm Weyl}\Biggl(
{\rm Riem}\Biggl[ \mathcal{Y}_{\alpha} \mathcal{Y}_{\gamma} \mathcal{Y}'_{\mu}
\mathcal{Y}'_{\rho}  \Bigl[ \mathcal{R}_{\beta\nu}
\mathcal{R}_{\delta\sigma} \!+\!  \mathcal{R}_{\beta\sigma}
\mathcal{R}_{\delta\nu}\Bigr] \Biggr] \Biggr) + O(\kappa^4). \qquad
\end{eqnarray}

A further simplification is to express the result
(\ref{finalanswer}) using covariant derivatives (with respect to the
de Sitter background) of the scalar propagator $i\Delta_A(x;x')$,
\begin{eqnarray}
\lefteqn{ \Bigl\langle \Omega \Bigl\vert
C_{\alpha\beta\gamma\delta}(x) \times C_{\mu\nu\rho\sigma}(x')
\Bigr\vert \Omega \Bigr\rangle = } \nonumber \\
& & \hspace{-.2cm} \frac{\kappa^2}{4} {\rm Weyl}\Biggl( {\rm
Riem}\Biggl[ D_{\alpha} D_{\gamma} D_{\mu}' D_{\rho}' i \Delta_A
\times \Bigl[ \mathcal{R}_{\beta\nu} \mathcal{R}_{\delta\sigma}
\!+\! \mathcal{R}_{\beta\sigma} \mathcal{R}_{\delta\nu}\Bigr]
\Biggr] \Biggr) + O(\kappa^4) \; . \qquad \label{altform}
\end{eqnarray}
(The flat space limit is obvious from this form.) The fact that the
three algebraicly independent tensor factors in expression
(\ref{finalanswer}) can be combined in this way is a consequence of
transversality (\ref{transverse}). Each of the three tensor factors
obeys all the algebraic symmetries of a product of two Weyl tensors,
but only a particular combination of all three obeys transversality.

Even more simplifications occur in $D = 4$ dimensions. For example,
the general form of $A''(y)$ from definition (\ref{Af}) contains an
infinite series,
\begin{eqnarray}
\lefteqn{A''(y) = \frac{H^{D-2}}{(4\pi)^{\frac{D}2}}\frac{1}{16}
\Biggl\{\Gamma\Bigl(\frac{D}{2} \!+\! 1\Bigr)
\Bigl(\frac{4}{y}\Bigr)^{ \frac{D}2 +1} \!+\! \Bigl(\frac{D}2 \!-\!
1\Bigr) \Gamma\Bigl(\frac{D}2 \!+\! 1\Bigr) \Bigl(\frac{4}{y}
\Bigr)^{\frac{D}2} } \nonumber \\
& & \hspace{-.2cm} + \sum_{n=1}^{\infty} \Biggl[ \frac{(n \!-\! 1)
\Gamma(n \!+\! D \!-\! 1)}{ \Gamma(n \!+\! \frac{D}{2})}
\Bigl(\frac{y}{4}\Bigr)^{n-2} \!\!- \frac{ (n \!-\! \frac{D}{2}
\!+\! 1 ) \Gamma(n \!+\! \frac{D}{2} \!+\! 1 )}{ (n \!+\! 1)!}
\Bigl(\frac{y}{4}\Bigr)^{n-\frac{D}{2}} \Biggr] \Biggr\} . \qquad
\label{Afpp}
\end{eqnarray}
However, only the first two terms survive for $D=4$,
\begin{equation}
\lim_{D \rightarrow 4} A''(y) = \frac{H^2}{16 \pi^2} \Biggl\{
\frac{8}{y^3} + \frac{2}{y^2} \Biggr\} \; .
\end{equation}

\subsection{Comparison with Previous Results}

\begin{table}

\vbox{\tabskip=0pt \offinterlineskip
\def\tablerule{\noalign{\hrule}}
\halign to200pt {\strut#& \vrule#\tabskip=1em plus2em& \hfil#\hfil&
\vrule#& \hfil#\hfil& \vrule#\tabskip=0pt\cr \tablerule
\omit&height4pt&\omit&&\omit&\cr \omit&height2pt&\omit&&\omit&\cr &&
$\!\!\!\! I \!\!\!\!$ && $\!\!\!\! D^{(I)} \!\!\!\!$ & \cr
\omit&height4pt&\omit&&\omit&\cr \tablerule
\omit&height2pt&\omit&&\omit&\cr && 1 &&
$-12\Big(\frac{4}{y}\Bigl)^3$ & \cr \omit&height2pt&\omit&&\omit&\cr
\tablerule \omit&height2pt&\omit&&\omit&\cr && 2 && $-18
(\frac4{y})^3 - 6 (\frac{4}{y})^2$ & \cr
\omit&height2pt&\omit&&\omit&\cr \tablerule
\omit&height2pt&\omit&&\omit&\cr && 3 &&
$6\Big(\frac{4}{y}\Bigl)^3+6\Big(\frac{4}{y}\Bigl)^2$ & \cr
\omit&height2pt&\omit&&\omit&\cr \tablerule
\omit&height2pt&\omit&&\omit&\cr && 4 &&
$-3\Big(\frac{4}{y}\Bigl)^3+3\Big(\frac{4}{y}\Bigl)^2$ & \cr
\omit&height2pt&\omit&&\omit&\cr \tablerule
\omit&height2pt&\omit&&\omit&\cr && 5 &&
$\frac{3}{2}\Big(\frac{4}{y}\Bigl)^3 +
\frac{3}{2}\Big(\frac{4}{y}\Bigl)^2$ & \cr
\omit&height2pt&\omit&&\omit&\cr \tablerule
\omit&height2pt&\omit&&\omit&\cr && 6 && $3\Big(\frac{4}{y}\Bigl)^2$
& \cr \omit&height2pt&\omit&&\omit&\cr \tablerule
\omit&height2pt&\omit&&\omit&\cr && 7 &&
$-\frac{1}{4}\Big(\frac{4}{y}\Bigl)^3 +
\frac{3}{4}\Big(\frac{4}{y}\Bigl)^2$ & \cr
\omit&height2pt&\omit&&\omit&\cr \tablerule}}

\caption{The coefficients $D^{(I)}$ of Kouris \cite{Kouris}
expressed in terms of our de Sitter length function $y(x;x')$. Each
term should be multiplied by $\frac{\kappa^2 H^6}{4\pi^2}$.}

\label{Kcoefs}

\end{table}

\begin{table}

\vbox{\tabskip=0pt \offinterlineskip
\def\tablerule{\noalign{\hrule}}
\halign to390pt {\strut#& \vrule#\tabskip=1em plus2em&
\hfil#\hfil& \vrule#& \hfil#\hfil& \vrule#\tabskip=0pt\cr
\tablerule
\omit&height4pt&\omit&&\omit&\cr
\omit&height2pt&\omit&&\omit&\cr
&& $\!\!\!\! I \!\!\!\!$
&& $\!\!\!\! S^{(I)}_{abcda'b'c'd'} \!\!\!\!$ & \cr
\omit&height4pt&\omit&&\omit&\cr
\tablerule
\omit&height2pt&\omit&&\omit&\cr
&& 1 && $\frac{1}{(4y-y^2)^2}\mathcal{Y}_a\mathcal{Y}_c
\mathcal{Y}'_{a'} \mathcal{Y}'_{c'} \Big[g_{bd}g_{b'd'} - 2
\mathcal{R}_{bb'} \mathcal{R}_{dd'} \Bigl]$ & \cr
\omit&height2pt&\omit&&\omit&\cr
\tablerule
\omit&height2pt&\omit&&\omit&\cr
&& 2 && $\frac{1}{4y-y^2} \mathcal{Y}_a \mathcal{Y}'_{c'} \mathcal{R}_{bb'}
\mathcal{R}_{cd'} \Big[\mathcal{R}_{da'} - \frac{1}{2(4-y)} \mathcal{Y}_{d}
\mathcal{Y}'_{a'} \Bigl]$ & \cr
\omit&height2pt&\omit&&\omit&\cr
\tablerule
\omit&height2pt&\omit&&\omit&\cr
&& 3 && $\frac{1}{4y-y^2} \mathcal{Y}_c \mathcal{Y}'_{c'} g_{bd}g_{a'd'}
\Big[ \mathcal{R}_{ab'} - \frac{1}{2(4-y)} \mathcal{Y}_{a} \mathcal{Y}'_{b'}
\Bigl]$ & \cr
\omit&height2pt&\omit&&\omit&\cr
\tablerule
\omit&height2pt&\omit&&\omit&\cr
&& 4 && $\frac{1}{4y-y^2} \Bigr[ g_{ac} \mathcal{Y}'_{a'} \mathcal{Y}'_{c'}
\mathcal{R}_{bd'} \mathcal{R}_{db'} + \mathcal{Y}_{a} \mathcal{Y}_{c}
g_{a'c'} \mathcal{R}_{bb'} \mathcal{R}_{dd'} \Bigl]$ & \cr
\omit&height2pt&\omit&&\omit&\cr
\omit&height2pt&\omit&&\omit&\cr
&& \omit && $-\frac{1}{2(4y-y^2)} \Bigl[ g_{ac} g_{bd} g_{b'd'}
\mathcal{Y}'_{a'} \mathcal{Y}'_{c'} + \mathcal{Y}_{a} \mathcal{Y}_{c}
g_{a'c'} g_{b'd'} g_{bd} \Bigr]$ & \cr
\omit&height2pt&\omit&&\omit&\cr
\tablerule
\omit&height2pt&\omit&&\omit&\cr
&& \omit && $\mathcal{R}_{ab'} \mathcal{R}_{bc'} \mathcal{R}_{cd'}
\mathcal{R}_{da'}$ & \cr
\omit&height2pt&\omit&&\omit&\cr
\omit&height2pt&\omit&&\omit&\cr
&& 5 && $-\frac{1}{2(4-y)} \Bigl[\mathcal{R}_{ab'} \mathcal{R}_{bc'}
\Bigl( \mathcal{R}_{cd'} \mathcal{Y}_{d} \mathcal{Y}'_{a'} +
\mathcal{R}_{da'} \mathcal{Y}_{c} \mathcal{Y}'_{d'} \Bigr) \Bigr]$ & \cr
\omit&height2pt&\omit&&\omit&\cr %%%
\omit&height2pt&\omit&&\omit&\cr
&& \omit  && $\hspace{2cm} + \mathcal{R}_{cd'} \mathcal{R}_{da'} \Bigl( 
\mathcal{R}_{ab'} \mathcal{Y}_{b} \mathcal{Y}'_{c'} + \mathcal{R}_{bc'}
\mathcal{Y}_{a} \mathcal{Y}'_{b'} \Bigr)\Bigr]$ & \cr
\omit&height2pt&\omit&&\omit&\cr 
\omit&height2pt&\omit&&\omit&\cr
&& \omit && $+\frac{1}{4(4-y)^2} \Bigl[ \mathcal{R}_{ab'}
\mathcal{R}_{cd'} \mathcal{Y}_{b} \mathcal{Y}_{d} \mathcal{Y}'_{a'}
\mathcal{Y}'_{c'} + \mathcal{R}_{bc'} \mathcal{R}_{da'} \mathcal{Y}_{a}
\mathcal{Y}_{c} \mathcal{Y}'_{b'} \mathcal{Y}'_{d'} \Bigr]$ & \cr
\omit&height2pt&\omit&&\omit&\cr
\tablerule
\omit&height2pt&\omit&&\omit&\cr
&& 6 && $g_{ac} g_{b'd'} \mathcal{R}_{da'} \mathcal{R}_{bc'}$ & \cr
\omit&height2pt&\omit&&\omit&\cr%%%
\omit&height2pt&\omit&&\omit&\cr && \omit && $+g_{ac} g_{b'd'}
\Bigl[-\frac{1}{2(4-y)} \Bigl( \mathcal{R}_{da'} \mathcal{Y}_b
\mathcal{Y}'_{c'} + \mathcal{R}_{bc'} \mathcal{Y}_d
\mathcal{Y}'_{a'} \Bigr) + \frac{1}{4(4-y)^2} \mathcal{Y}_b
\mathcal{Y}_d \mathcal{Y}'_{a'} \mathcal{Y}'_{c'} \Bigr]$ & \cr
\omit&height2pt&\omit&&\omit&\cr \tablerule
\omit&height2pt&\omit&&\omit&\cr && 7 && $g_{ac} g_{bd} g_{a'c'}
g_{b'd'}$ & \cr \omit&height2pt&\omit&&\omit&\cr \tablerule }}

\caption{The seed tensors $S^{(I)}_{abcda'b'c'd'}$ of Kouris
\cite{Kouris}, expressed using our standard basis tensors
(\ref{defs}-\ref{defs2}). Terms that drop after antisymmetrization
have been omitted.}.

\label{Kseeds}

\end{table}

In 2001 Kouris reported a result for the linearized Weyl-Weyl
correlator in $D=4$ dimensions \cite{Kouris}, derived using a de
Sitter invariant propagator in a general gauge \cite{INVPROP}.
Although the reader will recall from Section 1 that all these
propagators are illegitimate for one reason or another, the various
problems (spurious zero modes and invalid analytic continuations in
the constrained sector) should drop out of the Weyl-Weyl correlator. 
However, the Kouris result does not agree with ours, nor can his 
result be correct.

Kouris expressed his answer as a linear combination of scalar
functions (given in Table~\ref{Kcoefs}) times anti-symmetrized
tensor factors (the seeds for which are listed in
Table~\ref{Kseeds}),
\begin{equation}
\Bigl\langle \Omega \Bigl\vert C_{abcd}(x) \times C_{a'b'c'd'}(x')
\Bigr\vert \Omega \Bigr\rangle_{\rm Kouris} = \sum_{I=1}^7 D^{(I)}
\times S^{(I)}_{[ab] [cd] [a'b'] [c'd']} + O(\kappa^4) \; .
\label{Kanswer}
\end{equation}
The problem has to do with the various algebraic and differential
symmetries that the linearized Weyl-Weyl correlator must obey. We us
define,
\begin{equation}
W_{abcd a'b'c'd'} \equiv \sum_{I=1}^7 D^{(I)} \times S^{(I)}_{[ab]
[cd] [a'b'] [c'd']} \; ,
\end{equation}
This tensor should be, and is, anti-symmetric under interchange of
$(a,b)$, $(c,d)$, $(a',b')$ and $(c',d')$. However, it must also be
symmetric under the interchange of index pairs,
\begin{equation}
W_{abcda'b'c'd'} = W_{cdab a'b'c'd'} = W_{abcd c'd'a'b'} \; .
\label{prob1}
\end{equation}
Another symmetry inherited from the Riemann tensor is,
\begin{equation}
W_{a(bcd) a'b'c'd'} = 0 = W_{abcd a' (b'c'd')} \; . \label{prob2}
\end{equation}
The result must also be traceless within any index group. That is
obviously true on antisymmetric index pairs, but it must hold as
well for different pairs,
\begin{equation}
g^{ac} W_{abcd a'b'c'd'} = 0 = g^{a'c'} W_{abcd a'b'c'd'} \; .
\label{prob3}
\end{equation}
None of the algebraic symmetries (\ref{prob1}-\ref{prob3}) hold, nor
does the Kouris result obey transversality (\ref{transverse}),
\begin{equation}
D^a W_{abcd a'b'c'd'} = 0 = D^{a'} W_{abcd a'b'c'd'} \; .
\label{prob4}
\end{equation}

Kouris claimed to have checked (\ref{prob1}-\ref{prob2}) \cite{Kouris}. 
He does not seem to have realized that relations (\ref{prob3}) and
(\ref{prob4}) should hold. His choice of basis tensors is also
peculiar. There are 9 distinct invariant tensors with the algebraic
symmetries of two Riemann tensors --- antisymmetry plus relations
(\ref{prob1}-\ref{prob2}) \cite{TW7}. However, Kouris only used the
7 basis seeds listed in Table~\ref{Kseeds}. Enforcing tracelessness
(\ref{prob3}) should leave just three distinct tensors \cite{TW7},
and transversality (\ref{prob4}) should relate the coefficients of
these.

Of course our result (\ref{finalanswer}) obeys (\ref{prob1}-\ref{prob4})
so it cannot agree with (\ref{Kanswer}). It is not easy to compare the
two results termwise because Kouris employed the geometrical tensors 
(\ref{nun}-\ref{par}) of the mathematical physics convention.
However, it is simple enough to compare those terms which contain four
factors of $\mathcal{R}$. In our result (\ref{finalanswer}), with the
Kouris indices, these derive exclusively from the first term,
\begin{eqnarray}
\lefteqn{ \kappa^2 H^4 A''(y) \, {\rm Riem}\Biggl[ \Bigl[
\mathcal{R}_{a a'} \mathcal{R}_{c c'} \!+\! \mathcal{R}_{a c'}
\mathcal{R}_{c a'}\Bigr] \Bigl[\mathcal{R}_{b b'} \mathcal{R}_{d d'} \!+\! 
\mathcal{R}_{b d'} \mathcal{R}_{d b'}\Bigr] \Biggr] } \nonumber \\
& & \hspace{0cm} = -16 \kappa^2 H^4 A''(y) \Biggl( -\frac{1}{2 H^2}\Biggr)^4
\Biggl\{ 2 \frac{\partial^2 y}{\partial x^{a]} \partial {x'}^{[a'} }
\frac{\partial^2 y}{\partial {x'}^{b']} \partial x^{[c} }
\frac{\partial^2 y}{\partial x^{d]} \partial {x'}^{[c'} }
\frac{\partial^2 y}{\partial {x'}^{d']} \partial x^{[b} } \nonumber \\
& & \hspace{2cm} - \frac{\partial^2 y}{\partial x^{a} \partial {x'}^{[a'} }
\frac{\partial^2 y}{\partial {x'}^{b']} \partial x^{b} }
\frac{\partial^2 y}{\partial x^{c} \partial {x'}^{[c'} }
\frac{\partial^2 y}{\partial {x'}^{d']} \partial x^{d} } \nonumber \\
& & \hspace{4.5cm} - \frac{\partial^2 y}{\partial x^{a} \partial {x'}^{[c'} }
\frac{\partial^2 y}{\partial {x'}^{d']} \partial x^{b} }
\frac{\partial^2 y}{\partial x^{c} \partial {x'}^{[a'} }
\frac{\partial^2 y}{\partial {x'}^{b']} \partial x^{d} } 
\Biggr\} . \qquad \label{us}
\end{eqnarray}
The only one of Kouris's tensors which has four factors of $\mathcal{R}$
is $S^{(5)}_{abcda'b'c'd'}$. Note that in $D=4$ dimensions we can express
his $I = 5$ coefficient function in terms of $A''(y)$,
\begin{equation}
D^{(5)} = 48 \kappa^2 H^4 \times A''(y) \; .
\end{equation}
Retaining only the part of $S^{(5)}_{abcd a'b'c'd'}$ which contains
four factors of $\mathcal{R}$ gives,
\begin{eqnarray}
\lefteqn{ D^{(5)} \times S^{(5)}_{[ab] [cd] [a'b'c] [c'd']} } \nonumber \\
& & \hspace{-.4cm} \longrightarrow -48 \kappa^2 H^4 A''(y) \Biggl( -
\frac1{2 H^2}\Biggr)^4 
\frac{\partial^2 y}{\partial x^{a]} \partial {x'}^{[a'} }
\frac{\partial^2 y}{\partial {x'}^{b']} \partial x^{[c} }
\frac{\partial^2 y}{\partial x^{d]} \partial {x'}^{[c'} }
\frac{\partial^2 y}{\partial {x'}^{d']} \partial x^{[b} }
\; . \qquad \label{him}
\end{eqnarray}
Although the function of $y$ is tantalizingly close, the numerical
coefficients differ even between the parts of (\ref{us}) and (\ref{him})
which have the same tensor structure. One also sees the absence in
(\ref{him}) of the final two terms of (\ref{us}) which are needed to
enforce symmetries (\ref{prob1}-\ref{prob2}).

Two facts about Kouris's work make us suspect that it may be
resolved after correcting some minor errors:
\begin{itemize}
\item{The factors of $(4 - y)$ --- which are an artifact of the
cumbersome, de Sitter invariant notation --- all cancel in his final
result (\ref{Kanswer}); and}
\item{He claims to have checked relations (\ref{prob1}-\ref{prob2}),
even though they obviously fail for the result he reported.}
\end{itemize}
We accordingly explored minor emendations. One obvious possibility
is that Kouris's computer program might have generated the right
result, but he failed to extract the correct tensors owing to a
mistaken belief about the form the final answer should take. 
For example, his program might have generated all three of the 
$\mathcal{R}^4$ terms in (\ref{us}), but he might have imagined
that the result should include only the first one, and so checked
only its coefficient. (We guarded against this sort of error in our 
own analysis by reconstructing the form (\ref{NOg}) and checking 
that it really agrees with the long expression generated by our program.) 
However, we could not discover any way to make this work. 

Another possibility is that Kouris's antisymmetrized seed tensors 
would each obey (\ref{prob1}-\ref{prob2}) if they were first symmetrized 
with respect to the interchanges $(a,b) \leftrightarrow (c,d)$ and
$(a',b') \leftrightarrow (c',d')$. Perhaps he intended this, even though 
it was not stated? Unfortunately, this emendation still leaves a result 
which fails to obey (\ref{prob3}). That problem could be resolved by 
changing the sign of $D^{(6)}$ \cite{Atsushi}. However, the result still 
fails to obey (\ref{prob4}), and we are reluctant to consider more 
drastic emendations.

\subsection{Coincidence Limit}

Even had the result of Kouris been correct, it was unregulated by
virtue of being specialized to $D = 4$ dimensions. A simple but
powerful application of our formalism consists of taking the
coincidence limit of the Weyl-Weyl correlator using dimensional
regularization. To do this we set $a'=a$, $\Delta x^{\mu} = 0$ and 
$y = 0$. It is straightforward to read off the coincidence limit of 
each basis tensor from (\ref{defs}-\ref{defs2}) and 
(\ref{yb1})-(\ref{yb3}),
\begin{eqnarray}
\label{coin1} \displaystyle\lim_{x'\rightarrow x}
\mathcal{Y}_{\mu}(x;x') & = & \frac1{H}
\displaystyle\lim_{x'\rightarrow x}
\frac{\partial y}{\partial x^{\mu}} = 0 \; , \\
\displaystyle\lim_{x'\rightarrow x} \mathcal{Y}'_{\nu}(x;x') & = &
\frac1{H} \displaystyle\lim_{x'\rightarrow x}
\frac{\partial y}{\partial {x'}^{\nu}} = 0 \; , \\
\displaystyle\lim_{x'\rightarrow x} \mathcal{R}_{\mu\nu}(x;x') & = &
-\frac1{2 H^2} \displaystyle\lim_{x'\rightarrow x}
\frac{\partial^2y}{\partial x^{\mu} \partial {x'}^{\nu}} =
g_{\mu\nu}(x) \; .
\end{eqnarray}
Hence the seed tensors
$\sigma^{(2)}_{\alpha\beta\gamma\delta\mu\nu\rho\sigma}$ and
$\sigma^{(3)}_{\alpha\beta\gamma\delta\mu\nu\rho\sigma}$ both vanish
at coincidence and we have,
\begin{eqnarray}
\label{climit1} \lefteqn{\Bigl\langle \Omega \Bigl\vert
C_{\alpha\beta\gamma\delta}(x)
\times C_{\mu\nu\rho\sigma}(x) \Bigr\vert \Omega \Bigr\rangle } \nonumber \\
& & \hspace{1.5cm} = 4 \kappa^2 H^4 A''(0) \times {\rm Weyl}\Biggl(
{\rm Riem} \Bigl[ g_{\alpha (\mu} g_{\rho) \gamma} \, g_{\beta (\nu}
g_{\sigma) \delta} \Bigr] \Biggr) \!+\! O(\kappa^4) \; . \qquad
\end{eqnarray}
The coincidence limit of $A''(y)$ is also simple because we are
using dimensional regularization in which any $D$-dependent power of
zero vanishes. Hence only the $n=2$ term of the infinite series for
(\ref{Afpp}) survives,
\begin{equation}
\label{climit2} A''(0) = \frac{H^{D-2}}{(4\pi)^{\frac{D}2}} \times
\frac1{16} \frac{\Gamma(D \!+\! 1)}{\Gamma(\frac{D}2 \!+\! 2) } \; .
\end{equation}

Expanding the Weylized and Riemannized tensor factor in
(\ref{climit1}) gives,
\begin{eqnarray}
\label{climit3} \lefteqn{ {\rm Weyl}\Biggl( {\rm Riem} \Bigl[
g_{\alpha (\mu} g_{\rho) \gamma} \, g_{\beta (\nu} g_{\sigma)
\delta} \Bigr] \Biggr) = 4 g_{\alpha [\mu} g_{\nu ] \beta} \,
g_{\gamma [\rho} g_{\sigma ] \delta} + 4 g_{\alpha [\rho} g_{\sigma
]\beta} \, g_{\gamma [\mu} g_{\nu ] \delta} }
\nonumber \\
& & - 8 g_{\alpha ] [\mu} g_{\nu ] [\gamma} g_{\delta ] [\rho}
g_{\sigma ] [\beta} + \frac{24}{D \!-\! 2} \Biggl( g_{\alpha ]
[\gamma} g_{\delta ] [\mu} g_{\nu ] [\rho} g_{\sigma ] [\beta} +
g_{\alpha ] [\gamma}
g_{\delta] [\rho} g_{\sigma] [\mu} g_{\nu] [\beta} \Biggr) \nonumber \\
& & \hspace{6cm} + \frac{24}{(D \!-\! 2) (D \!-\! 1)} \, g_{\alpha
[\gamma} g_{\delta] \beta} \, g_{\mu [\rho} g_{\sigma ] \nu} \; .
\qquad
\end{eqnarray}
What we are ultimately interested in is the coincident Weyl-Weyl
correlator with the indices properly contracted. That is, we
contract $g^{\alpha\mu}g^{\beta\nu}g^{\gamma\rho}g^{\delta\sigma}$
into (\ref{climit3}) to obtain,
\begin{equation}
\label{Csquared} \Bigl\langle \Omega \Bigl\vert
C^{\alpha\beta\gamma\delta}(x) C_{\alpha\beta\gamma\delta}(x)
\Bigr\vert \Omega \Bigr\rangle = 4 (D\!-\!3) D (D \!+\! 1) (D\!+\!
2) A''(0) \kappa^2 H^4 + O(\kappa^4 H^8) \; .
\end{equation}

\section{Discussion}

Our result for the linearized Weyl-Weyl correlator is
(\ref{finalanswer}). It does not agree with what Kouris obtained
\cite{Kouris}, but that result cannot be correct because it lacks
some of the algebraic symmetries of the Weyl tensor and is not
transverse. By taking the coincidence limit of our result (with
dimensional regularization) and contracting the indices we derived
an expression (\ref{Csquared}) for the expectation value of
$C^{\alpha\beta\gamma\delta}(x) C_{\alpha\beta\gamma\delta}(x)$ at
lowest order.

Despite the fact that our propagator shows a physical breaking of de
Sitter invariance \cite{Kleppe}, the Weyl-Weyl correlator computed
from it is completely de Sitter invariant at linearized order. There
are different opinions about why this happened. Mathematical
physicists maintain that it is because ``free gravitons'' are de
Sitter invariant. They hold that the de Sitter breaking manifest in
our propagator is merely a gauge artifact which drops out when
linearized gauge invariance is enforced by going to the linearized
Weyl-Weyl correlator \cite{FH,HMM}. We do not agree. We believe the
de Sitter breaking terms dropped out because the logarithmic
infrared divergence from which they derive is rendered convergent
(and hence de Sitter invariant) by the derivatives needed to convert
the graviton field into a linearized Weyl tensor. This was so
obvious that it was noted even before the computation was begun
\cite{MTW1}.

At this point we should comment on what one learns about gravity
from the linearized Weyl-Weyl correlator versus the undifferentiated
propagator. The dynamical variable of gravity is the metric and,
like all local force fields, it consists of three things:
\begin{itemize}
\item{A pure gauge part which fixes how we measure
lengths and times;}
\item{A constrained part which carries the gravitational response to
sources of stress-energy; and}
\item{A dynamical part which represents gravitational radiation.}
\end{itemize}
In a gauge such ours \cite{TW1,RPW1}, the graviton propagator
contains all three of these things. By insisting on the linearized
Weyl tensor in order to expunge the pure gauge part, mathematical
physicists have edited out the constrained fields and they have also
weighted infrared graviton modes much less strongly than ultraviolet
ones. There is no question that this abandons perfectly physical and
gauge invariant information. For example, the constrained part of
the gauge fixed propagator provides the gravitational response to
matter, which comprises all but one of the classic tests of general
relativity. And the canonical weighting of graviton modes is
reflected in the scale invariance of the tensor power spectrum
(\ref{Deltah}).

It seems clear to us that this controversy over the relevance of the
gauge fixed graviton versus the linearized Weyl tensor is identical
to one which was finally settled for electromagnetism by the
Aharonov-Bohm effect \cite{bohm}. It is a gauge invariant fact that
matter fields couple to the electromagnetic vector potential, not to
the field strength. This implies that the undifferentiated vector
potential is itself observable in a fixed gauge. Similarly, it is a
gauge invariant fact that matter --- and even gravity itself
--- couples to the undifferentiated graviton field, not to the
curvature. The same reasoning implies that the undifferentiated
graviton field must be observable in a fixed gauge. Indeed,
strenuous efforts \cite{PLANCK,EBEX,SPIDER,PIPER} are underway to
measure the tensor power spectrum (\ref{Deltah}) which is precisely
such an observable. Concerns over invariance should be resolved in
gravity the very same way as in gauge theories: by noting that a
quantity can always be defined invariantly by specifying it in a
fixed gauge. (For examples, see \cite{TW8,RPW2}.)

An interesting parallel exists with the free massless, minimally
coupled scalar on a non-dynamical de Sitter background,
\begin{equation}
\mathcal{L} = -\frac12 \partial_{\mu} \varphi \partial_{\nu} \varphi
g^{\mu\nu} \sqrt{-g} \; .
\end{equation}
There is no question that this theory breaks de Sitter invariance
\cite{AF,VFLS}. If one defines things so as to preserve the
homogeneity and isotropy of cosmology then the scalar propagator is
precisely the same as the spatial polarizations of our graviton
field \cite{OW},
\begin{equation}
\Bigl\langle \Omega \Bigl\vert T\Bigl[ \varphi(x) \varphi(x')\Bigr]
\bigr\vert \Omega \Bigr\rangle = i\Delta_A(x;x') = A\Bigl(
y(x;x')\Bigr) + k \ln(a a') \; . \label{phiprop}
\end{equation}
However, because all fields in the stress tensor are differentiated,
the expectation value of the free scalar stress tensor happens to be
de Sitter invariant \cite{AF},
\begin{equation}
\Bigl\langle \Omega \Bigl\vert T_{\mu\nu} \Bigr\vert \Omega
\Bigr\rangle = \Bigl( \delta^{\rho}_{\mu} \delta^{\sigma}_{\nu}
\!-\! \frac12 g_{\mu\nu} g^{\rho\sigma} \Bigr) \lim_{x' \rightarrow
x} \partial_{\rho} \partial_{\sigma}' i\Delta_A(x;x') = (D \!-\! 2)
H^2 A'(0) g_{\mu\nu} \; .
\end{equation}
People who believe passionately in de Sitter invariance have been
known to proclaim this result as evidence that the de Sitter
breaking of the scalar propagator (\ref{phiprop}) is ``unphysical.''
However, it is nothing more nor less than the result of the de
Sitter breaking infrared divergence being logarithmic, so that
derivatives eliminate it.

Now add an interaction which involves undifferentiated scalars,
\begin{equation}
\mathcal{L} = -\frac12 \partial_{\mu} \varphi \partial_{\nu} \varphi
g^{\mu\nu} \sqrt{-g} - \frac{\lambda}{4!} \varphi^4 \sqrt{-g} + {\rm
Counterterms} \; .
\end{equation}
Because the interacting theory contains undifferentiated scalars,
the expectation value of the stress tensor shows explicit de Sitter
breaking \cite{OW,KOW},
\begin{eqnarray}
\lefteqn{\Bigl\langle \Omega \Bigl\vert T_{\mu\nu} \Bigr\vert \Omega
\Bigr\rangle = (D \!-\! 2) H^2 A'(0) g_{\mu\nu} } \nonumber \\
& & \hspace{.5cm} - \frac{\lambda H^4}{(4 \pi)^4} \Biggl\{ \Bigl[ 2
\ln^2(a) \!+\! \frac72 \ln(a) \Bigr] g_{\mu\nu} + \Bigl[ \frac43
\ln(a) \!+\! \frac{13}{18} \Bigr] \mathcal{T}_{\mu}
\mathcal{T}_{\nu} \Biggr\} + O(\lambda^2) \; . \qquad
\end{eqnarray}
de Sitter breaking has also been exhibited for the
one-particle-irreducible (1PI) 2-point function at one and two loop
orders \cite{BOW}, and one can show generally that each additional
power of $\lambda$ in a 1PI function produces up to two additional
de Sitter breaking factors of $\ln(a)$ \cite{TW9}.

The same sort of de Sitter breaking goes on {\it whenever} one adds
interactions which involve undifferentiated scalars on non-dynamical
de Sitter background. Explicit, fully renormalized results exists at
one and two loop orders for scalar quantum electrodynamics
\cite{SQED} --- which shows one factor of $\ln(a)$ for each factor
of the loop counting parameter $e^2$ --- and for Yukawa theory
\cite{Yukawa} --- which shows one factor of $\ln(a)$ for each
additional loop. Similar results have even been obtained for the
nonlinear sigma model \cite{KK}.

Let us now take note of the undifferentiated graviton interactions
which abound in the gravitational Lagrangian (\ref{Linv}). Based on
the known relation between interactions and infrared logarithms, one
expects an additional factor of $\ln(a)$ for each extra factor of
the quantum gravitational loop counting parameter $\kappa^2$
\cite{TW9}. Which brings us to the observation that $\langle \Omega
\vert C^{\alpha\beta\gamma\delta}(x) C_{\alpha\beta\gamma\delta}(x)
\vert \Omega \rangle$ can show de Sitter breaking at order
$\kappa^4$. Individual diagrams certainly make such contributions,
but it might be that they all add up to zero. We propose that this
be checked.

It should be noted that the operator $C^{\alpha\beta\gamma\delta}(x)
C_{\alpha\beta\gamma\delta}(x)$ is a scalar, rather than a true
invariant. Promoting it to an invariant requires somehow fixing the
observation point $x^{\mu}$, and that would inevitably involve
nonlocality. However, the expectation value of
$C^{\alpha\beta\gamma\delta}(x) C_{\alpha\beta\gamma\delta}(x)$
should serve as a test of the physical de Sitter invariance of the
gauge fixed theory. And this quantity has a priceless advantage over
invariant (and hence nonlocal) observables: {\it we know how to
renormalize it.}

\vskip .5cm

\centerline{\bf Acknowledgements}

We thank A. Higuchi for correspondence concerning the result of
his former student S. Kouris. This work was partially supported
by NSF grant PHY-0855021 and by the Institute for Fundamental
Theory at the University of Florida.

\end{document}